\documentclass[aps,prx,reprint,twocolumn,amsmath,amssymb,amsfonts,superscriptaddress]{revtex4-1}
 
\usepackage{tikz}
\usepackage{tikz-cd}
\usepackage{amsmath}
\usepackage{hyperref}
\usepackage{color}
\usepackage{multirow}
\usepackage[nolist,nohyperlinks]{acronym}  
\usepackage{microtype}

\usetikzlibrary{positioning}

\usepackage{times}
\usepackage[varg]{txfonts}
 
\usepackage{mathrsfs}
\usepackage{color}
\usepackage{graphicx}
\usepackage[percent]{overpic}
\usepackage{bm}
\usepackage{natbib}

\usepackage{nicefrac}

\newcommand{\termsym}[3]{{}^{#1}{#2}_{#3}}

\newcommand{\bra}[1]{\left \langle #1\right|} 
 
 \newcommand{\ket}[1]{\left| #1  \right \rangle}
\newcommand{\braket}[2]{\left \langle  #1 | #2 \right \rangle}

\newcommand{\hc}{{\rm h.c.}}
\newcommand{\avg}[1]{\langle #1 \rangle}

\newcommand{\h}[1]{{#1}^{\dagger}} 
\newcommand{\trp}[1]{{#1}^{\intercal}} 
\newcommand{\cc}[1]{{#1}^{*}}\newcommand{\cb}[1]{\bar{#1}}

\newcommand{\sgn}{{\rm sgn}}

\newcommand{\meV}{\ {\rm meV}}

\newcommand{\K}{\ {\rm K}}
\newcommand{\mK}{\ {\rm mK}}

\newcommand{\tsup}[1]{\textsuperscript{#1}}
\newcommand{\tsub}[1]{\textsubscript{#1}}

\newcommand{\rth}[1]{{#1}\tsup{3+}}

\newcommand{\abo}[2]{#1\tsub{2}#2\tsub{2}O\tsub{7}}
\newcommand{\eto}{\abo{Er}{Ti}}

\newcommand{\yto}{\abo{Yb}{Ti}}

\newcommand{\spinel}[3]{{#1}{#2}\tsub{2}{#3}\tsub{4}}
\newcommand{\byzo}{Ba\tsub{3}Yb\tsub{2}Zn\tsub{5}O\tsub{11}}
\newcommand{\ybmggao}{YbMgGaO\tsub{4}}

\renewcommand{\vec}[1]{\boldsymbol{#1}}
\newcommand{\mat}[1]{\vec{#1}}
\newcommand{\vhat}[1]{\vec{\hat{#1}}}

\newcommand{\Uf}{U^+_f}
\newcommand{\DeltaOx}{\Delta}

\newcommand{\f}{f}

\definecolor{cred}{RGB}{228,26,28}
\definecolor{cblue}{RGB}{55,126,184}
\definecolor{cdblue}{RGB}{40,96,139}
\definecolor{clblue}{RGB}{205,223,237}
\definecolor{cgreen}{RGB}{77,175,74}
\definecolor{cgray}{RGB}{150,150,150}
\definecolor{clgray}{RGB}{200,200,200}
\definecolor{cpurple}{RGB}{152,78,163}
\definecolor{corange}{RGB}{255,127,0}
\definecolor{cgold}{RGB}{230,171,2}

\definecolor{ccc}{RGB}{40,96,139}

\hypersetup{colorlinks=true,linkcolor=ccc,citecolor=ccc,urlcolor=ccc} 

\newacro{DM}[DM]{{Dzyaloshinskii-Moriya}}
\newacro{AIAO}[AIAO]{all-in/all-out}
\newacro{PC}[PC]{Palmer-Chalker}
\newacro{SFM}[SFM]{splayed ferromagnet}
 
\begin{document}
 
\title{Frustration and anisotropic exchange in ytterbium magnets with edge-shared octahedra}
\author{Jeffrey G. Rau}
\affiliation{Department of Physics and Astronomy, University of
  Waterloo, Ontario, N2L 3G1, Canada}
\affiliation{Max-Planck-Institut f\"ur Physik komplexer Systeme, 01187 Dresden, Germany}
\author{Michel J. P. Gingras} 
\affiliation{Department of Physics and Astronomy,
University of Waterloo, Ontario, N2L 3G1, Canada}
\affiliation{Perimeter Institute for Theoretical Physics, Waterloo,
Ontario, N2L 2Y5, Canada} 
\affiliation{Canadian Institute for Advanced
Research, 180 Dundas Street West, Suite 1400, Toronto, ON, M5G 1Z8,
Canada}
\date{\today}

\begin{abstract}
We consider the structure of anisotropic exchange interactions in ytterbium-based insulating rare-earth magnets built from edge-sharing octahedra. We argue the features of trivalent ytterbium and this structural configuration allow for a qualitative determination of the different anisotropic exchange regimes that may manifest in such compounds. The validity of such super-exchange calculations is tested through comparison to the well-characterized breathing pyrochlore compound \byzo{}. With this in hand, we then consider applications to three-dimensional pyrochlore spinels as well as two-dimensional honeycomb and triangular lattice systems built from such edge-sharing octahedra. We find an extended regime of robust emergent weak anisotropy with dominant antiferromagnetic Heisenberg interactions as well as smaller regions with strong anisotropy. We discuss the implications of our results for known compounds with the above structures, such as the spinels \spinel{A}{Yb}{X} ($A$ = Cd, Mg, $X$ = S, Se), the triangular compound \ybmggao{}, which have recently emerged as promising candidates for observing unconventional magnetic phenomena. Finally, we speculate on implications for the \abo{R}{M} pyrochlore compounds and some little studied honeycomb ytterbium magnets.
\end{abstract}

\maketitle

\section{Introduction}
Frustration generated by anisotropic exchange interactions has attracted a significant amount of attention recently~\cite{krempa-2014-arcmp,rau-2016-arcmp}. Induced by strong spin-orbit coupling~\cite{rau-2016-arcmp,trebst2017kitaev,hermanns-review}, this kind of frustration is distinct from the usual geometric type~\cite{balents2010spin}, as it does not solely rely on the structure of the underlying crystal lattice. Instead, different types of competing anisotropic exchange interactions compatible with the discrete symmetries of the crystal need to be tuned to induce strong frustration. 

A noteworthy example of this kind of physics is the recent development of ``Kitaev magnetism'' in Mott insulators with strong spin-orbit coupling~\cite{krempa-2014-arcmp,rau-2016-arcmp}. The canonical example of this physics is found in iridium oxides~\cite{jackeli-2009-mott}, or iridates, where the relevant atomic states are a spin-orbital mixed $J_{\rm eff}=1/2$ doublet~\cite{kim-2008-spin-orbit}. As pointed out in the pioneering work of \citet{jackeli-2009-mott}, by building such an iridate out of edge-sharing IrO\tsub{6} octahedra, one can realize dominant bond-dependent Ising interactions~\cite{kitaev-2006-anyons}. If arranged in a honeycomb~\cite{kitaev-2006-anyons} or honeycomb-like~\cite{mandal2009exactly} lattice, this bond-dependent interaction realizes the exactly solvable spin-1/2 model first studied by \citet{kitaev-2006-anyons}. This model has a number of intriguing features but, foremost, has attracted significant attention~\cite{rau-2016-arcmp,winter-review,hermanns-review} as it hosts a concrete example of a $Z_2$ spin liquid with its associated fractionalized excitations~\cite{kitaev-2006-anyons}. These systems represent an exchange \emph{regime} distinct from the usual Heisenberg, Ising or XY type models, one that \emph{only} appears in the limit of very strong spin-orbit coupling.

In this article, we explore this anisotropic exchange physics from a somewhat different perspective: what kind of well-defined (anisotropic) exchange regimes can we find in \emph{rare-earth} magnets? We argue that such regimes \emph{do} exist and, further, that they can shed light on the physics of known ytterbium based magnets as well as suggest promising new materials to explore. There are many inherent advantages of rare-earths over transition metal magnets: the most prominent of these is in being much stronger Mott insulators as well as having very large spin-orbit coupling. The issues of direct orbital overlap and of further neighbor interactions (that can complicate some transition metal magnets) are thus strongly suppressed, as are any notions of itinerant or ``weak'' Mott insulator behavior. Further, given the somewhat uniform chemistry of the rare-earths, substitution of one rare-earth for another is significantly less disruptive than in transition metal compounds. This affords numerous opportunities in synthesizing new compounds, as well as doping or diluting a given compound to probe its physics.  However, there are also downsides, most notably the much smaller energy scales inherent to rare-earth magnets.  Typically, one should expect exchange interactions to be roughly two orders of magnitude smaller than in transition metal magnets with similar inter-atomic distances.  Theoretically, there is also the issue of the more complex atomic and super-exchange~\cite{onoda-2011-quantum,petit2014,rau-2015-magnitude,iwahara-2015-exchange} physics in rare-earths.  This, combined with the small energy scales, can make reliable determination of the exchange interactions difficult. To circumvent these difficulties, one usually relies on extracting the possible symmetry allowed exchange parameters through direct fitting to some manageable experimental limit, such as through high-temperature expansions~\cite{applegate2012,hayre2013,oitmaa2013obd,jaubert2015multiphase}, spin-wave spectra in high magnetic fields~\cite{ross-2011-quantum,savary2012obd,petit2015dynamics,thompson2017quasiparticle} or through various local probes~\cite{li-2015-rare-earth}. Even with extensive data and a controlled theoretical regime, this approach can still fail to determine the exchange parameters uniquely~\cite{petit2015dynamics,thompson2017quasiparticle}.

We argue in this paper that some of the theoretical complications discussed above are absent in ytterbium-based rare-earth magnets which have the same edge-sharing structure that is found in Kitaev materials. First, due to the nearly filled $\f^{13}$ electronic configuration, the atomic states are less complex than in a typical rare-earth ion.  The single low-lying $\f^{14}$ state and the (relative) simplicity of the higher lying $\f^{12}$ states enables some simplifications of the computation of the magnetic interactions. Second, the crystal field energy scale in ytterbium compounds is typically large, yielding little effect from virtual crystal field excitations~\cite{molavian2007,petit2014,rau-2016-order}. Third, the equivalence of the two ligand exchange paths in this edge-shared configuration leads to fewer orbital overlap parameters than in cases with two inequivalent exchange paths, such as in the \abo{R}{M} pyrochlores~\cite{onoda-2011-quantum}. Fourth, due to the low angular momentum ($J=7/2$), one expects that, irrespective of the composition of the crystal field ground doublet, the interactions between \rth{Yb} ions are generically quantum~\cite{rau-2015-magnitude}, with all of the symmetry allowed exchanges being potentially significant. Rather importantly, this edge-sharing structure is realized in many material contexts: most notably in rare-earth pyrochlore chalcogenide spinels~\cite{lau2005spinel,higo-2016-frustrated} of the form \spinel{A}{Yb}{X} where A = Cd, Mg and X = S, Se. It also appears in the recent spin liquid candidate \ybmggao{}~\cite{li-2015-rare-earth} as well as potentially in several, heretofore little studied, honeycomb compounds such as YbCl\tsub{3}~\cite{templeton1954crystal}.  A theoretical approach to explore possible regimes of anisotropic exchange in a wide range of materials with a diverse set of lattice geometries would be therefore appear to be broadly useful.

Our primary goal is calculating the exchanges in such ytterbium based magnets.  Given the approximate nature of these calculations, it is important that we can validate our approach. Thankfully, we are furnished with an excellent test case for this line of attack: the recently well-characterized ``breathing'' pyrochlore compound \byzo{}~\cite{kimura-2014-byzo,haku-2016-cef}. In this compound, the full anisotropic exchange interactions can be determined quite precisely~\cite{rau-2016-byzo,haku-2016-byzo,park-2016-byzo} through a direct comparison to thermodynamic and inelastic neutron scattering data, thanks to its nature as a few-body problem. Somewhat surprisingly, these interactions carry significant structure; there is a dominant antiferromagnetic Heisenberg exchange, large \ac{DM} interaction and very small symmetric anisotropies~\cite{rau-2016-byzo,haku-2016-byzo,park-2016-byzo}. We show that a direct calculation of the super-exchange processes can capture precisely this physics: by tuning the composition of the crystalline electric field ground doublet, there exists a robust regime of parameter space with these precise characteristics.

Emboldened by this agreement, we then consider these calculations for idealizations of the pyrochlore spinels, of the triangular lattice compounds and for the little studied honeycomb rare-earth magnets. We find that there are well-defined regimes in these parameter spaces; that is, limits where certain interaction channels \emph{strongly} dominate over the others. Most prominently, we find a robust regime with strong antiferromagnetic Heisenberg interactions and sub-dominant \ac{DM} interactions, as found for \byzo{}. This is analogous to the kind of emergent, ``weak'' anisotropy that can appear in transition metal oxides with strong spin-orbit coupling~\cite{rau-2016-arcmp} formed of corner-sharing octahedra. We further find smaller, more fragile regions with dominant Kitaev and other anisotropic exchange interactions. Our work is thus a ``proof of principle" that intrinsically anisotropic exchange regimes, such as that found in transition metal Kitaev materials, can also be found in rare-earth magnets. Further, these calculations provide a concrete example of how single-ion and exchange anisotropies can be independent when spin-orbit coupling is strong.  We further consider the robustness of our choice of microscopic parameters, specifically the Slater-Koster overlap parameters and the atomic energy scales, arguing that they do not qualitatively affect much of our results. 

Next, we discuss applications of our results to real materials. In the \spinel{A}{Yb}{X} spinels, where the crystal field parameters can be estimated, we compute the exchange interactions and speculate on possible implications for their physics. In particular, we find that the weak anisotropy regime found in \byzo{} also appears in these compounds with a dominant antiferromagnetic Heisenberg exchange and subdominant indirect \ac{DM} interaction~\cite{canals2008}. For the full, non-breathing pyrochlore lattice, this is a classical phase boundary between two ordered states, a ferromagnet and an antiferromagnet with an accidental $U(1)$ degeneracy. We find that the small symmetric anisotropies push the spinels into the antiferromagnetic phase with the $U(1)$ degeneracy resolved by quantum order-by-disorder. Finally, we argue that this parameter regime and its proximity to this phase boundary has a direct analogue in the pyrochlore \abo{Yb}{M} family (where M = Ti, Ge, Sn), and that the spinels may exhibit the same unusual dynamics found in these compounds~\cite{hallas2016universal}.  For \ybmggao{}, we investigate the possible effects of Mg/Ga disorder on the exchanges, both through changes in the crystal field and in the ligand bond angles Finally, we provide some outlook what one may learn more broadly from these calculations about ytterbium-based magnets; explicitly, from the fact that there exist these well-defined limits \emph{at all} in what would have na\"ively been expected to be a somewhat arbitrary parametrization. We also identify a region in parameter space that is highly sensitive to the details of the atomic physics and ligand environment. We argue this sensitivity may be relevant to the exchange physics in ytterbium pyrochlores of the form \abo{Yb}{M}. We hope the possibilities suggested here may lead to further work to characterize rare-earth magnetism and other unusual exchange regimes on a wider variety of material contexts and to discover new interesting states of matter in insulating ytterbium-based magnets.

This article is structured as follows: in Sec.~\ref{sec:single-ion}, we give an overview of the atomic physics of \rth{Yb}. In Sec. \ref{sec:two-ion}, we introduce the relevant symmetry allowed anisotropic exchange models, before discussing the super-exchange processes relevant for edge-shared Yb ions in Sec.~\ref{sec:super-exchange}. The qualitative correctness of this methodology is benchmarked for \byzo{} in Sec.~\ref{sec:validation}. We then explore the cubic crystal field limits as detailed in Sec.~\ref{sec:cubic-limit}, before discussing the general crystal field results in Sec.~\ref{sec:general}.  We distinguish two cases: those without local frames for the crystal fields, such as the triangular and honeycomb lattices and those with local frames, such as the spinel and breathing pyrochlore lattices. With these results in hand. we discuss  applications to real materials in Sec.~\ref{sec:applications}, specifically the pyrochlore spinels \spinel{A}{Yb}{X} (Sec.~\ref{sec:spinels}) and the triangular compound \ybmggao{} (Sec.~\ref{sec:triangular}). Finally, in Sec.~\ref{sec:discussion}, we hypothesize on the properties of possible rare-earth magnets that may realize the honeycomb (or hyper-honeycomb) structure and present a more general outlook for rare-earth magnetism.

\begin{figure*}[!htpb]
  \centering
  \begin{overpic}[width=0.6\columnwidth]{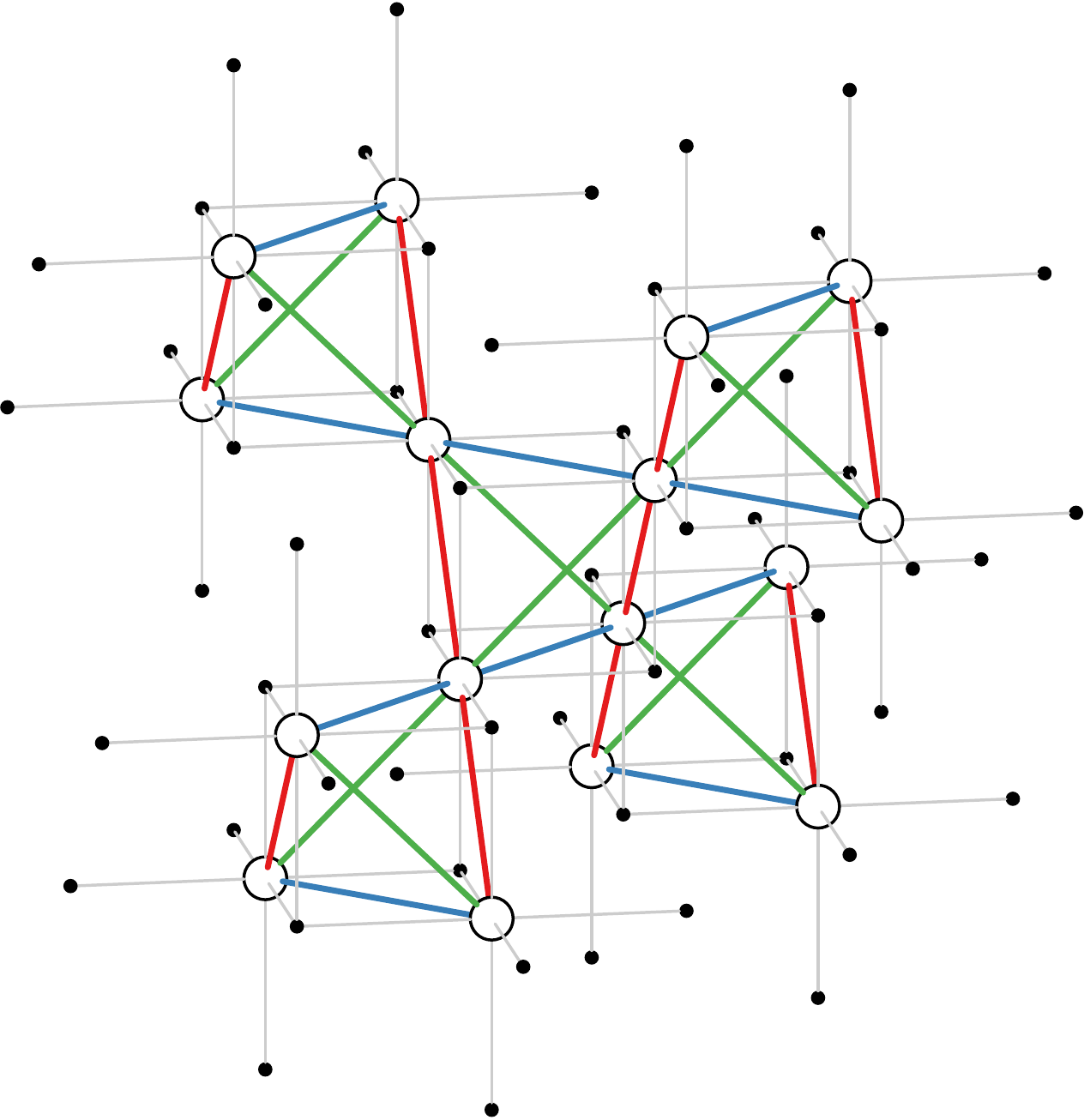}
  \put(20,30){\textcolor{cred}{$x$}}
  \put(32,23){\textcolor{cgreen}{$y$}}
  \put(30,38){\textcolor{cblue}{$z$}}
  \end{overpic}
  \centering
  \begin{overpic}[width=0.65\columnwidth]{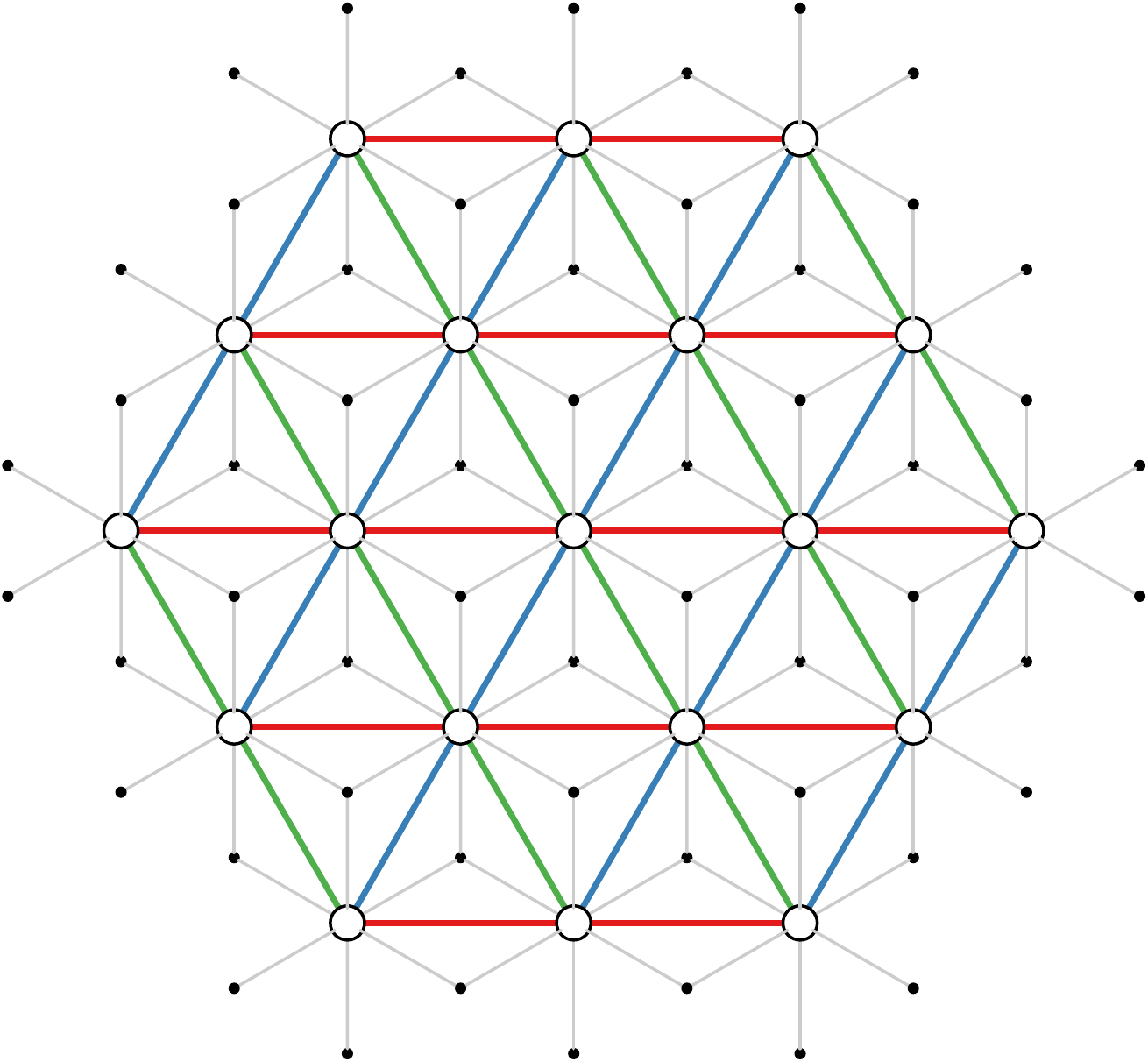}
  \put(28.5,31){\textcolor{cred}{$x$}}
  \put(21,20){\textcolor{cgreen}{$y$}}
  \put(37,20){\textcolor{cblue}{$z$}}
  \end{overpic} 
  \centering
  \begin{overpic}[width=0.6\columnwidth]{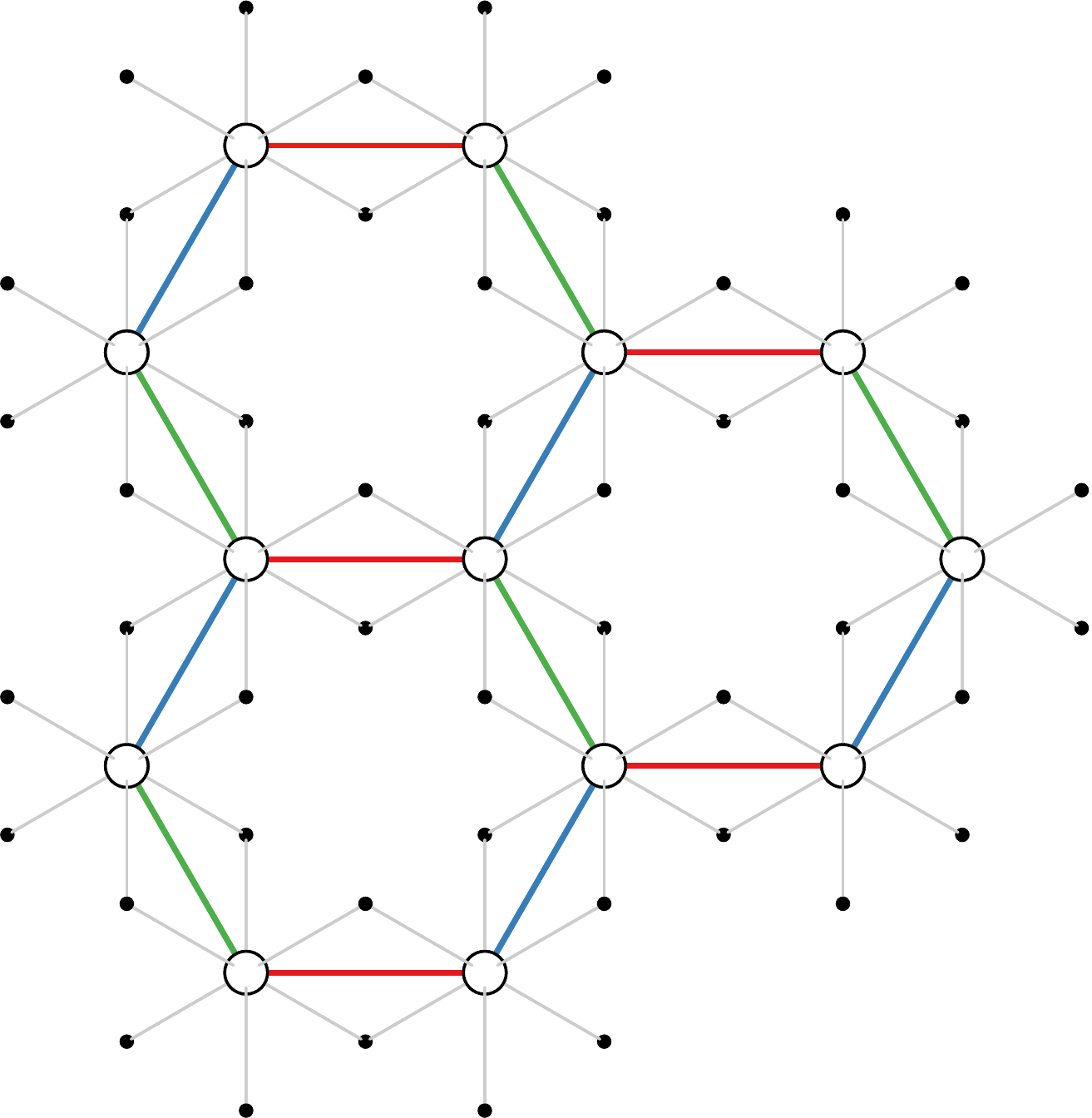}
  \put(31,52){\textcolor{cred}{$x$}}
  \put(12,58){\textcolor{cgreen}{$y$}}
  \put(12,40){\textcolor{cblue}{$z$}}
  \end{overpic} 
  \caption{
    \label{fig:structures}
    Crystal structures built from edge-shared octahedra. We show (a)
pyrochlore, (b) triangular and (c) honeycomb structures.  The ligands
sit between nearest-neighbor sites forming $90^\circ$ bonds for the
ideal case shown. For each lattice type, the three symmetry related bond types are denoted as $x$,
$y$ and $z$, shown in red, green and blue.
  }
\end{figure*}
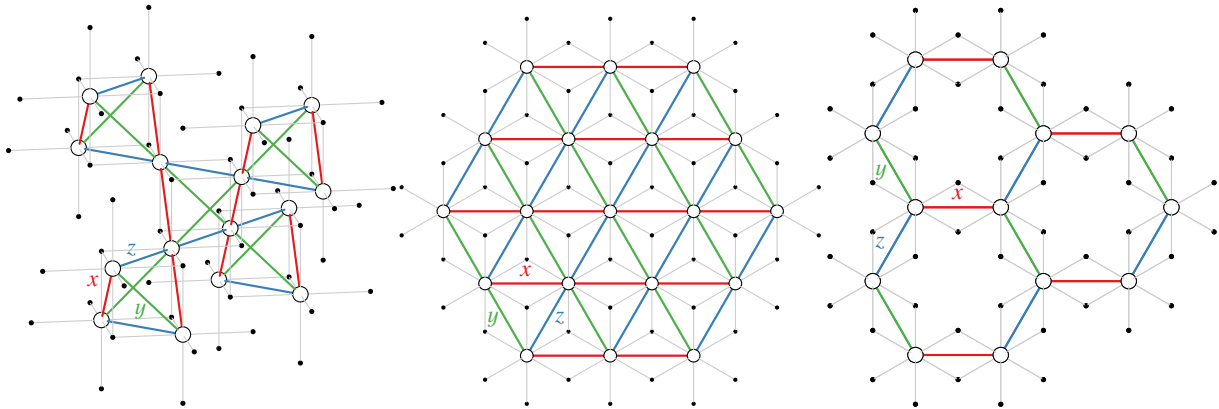

\section{Single-ion physics}
\label{sec:single-ion}

We begin by determining the effective interactions between \rth{Yb} ions in the edge-sharing structures discussed in the Introduction. These kinds of lattices have been studied in detail in the context of transition metal spin-orbit Mott insulators, such as iridium oxides~\cite{rau-2016-arcmp}. There are large number of structures one can form this way; this include honeycomb, triangular, pyrochlore lattices (illustrated in Fig.~\ref{fig:structures}) as well as more baroque lattices such as hyper-honeycomb~\cite{hypertakagi}, the harmonic-honeycomb series~\cite{modic2014new}, the hyper-octagon~\cite{hermanns2014} and the hyper-kagome structures~\cite{okamoto2007hyper}.  We refer the reader to Refs.~[\onlinecite{wells1977three},\onlinecite{obrien2016}] for a more complete catalog of these lattices. In all cases of interest, we consider the common \rth{Yb} valence.

Since \rth{Yb} has an $\f^{13}$ electronic configuration, we can consider only the low-lying ${}^2 F_{7/2}$ multiplet with $J=7/2$, $L=3$ and $S=1/2$. The eight-fold degeneracy of these levels is lifted in a crystal environment.  In the compounds of interest, the \rth{Yb} ion is surrounded by an approximately octahedral cage of ligands. Na\"ively, we may then expect the dominant contributions to the crystalline electric field to have full cubic symmetry. If this is so, the ${}^2F_{7/2}$ states split into two doublets of types $\Gamma_6$ and $\Gamma_7$ and a quartet of type $\Gamma_8$~\cite{lea1962raising}. Typically for the kind of octahedral cage of interest here, both experimentally and theoretically~\cite{lea1962raising}, the ground state is of type $\Gamma_6$ and is separated from the other two states by a large energy gap, of order $\sim 30-50 \meV$ ~\cite{spinel-cef-1,spinel-cef-2,higo-2016-frustrated,haku-2016-cef,paddison_continuous_2016}. For example, in the ytterbium spinels one obtains values for this gap of order $\sim 20 \meV$ ~\cite{spinel-cef-1,spinel-cef-2,higo-2016-frustrated}. In the breathing pyrochlore \byzo{}, one finds a gap of $\sim 38 \meV$~\cite{haku-2016-cef} and in the triangular compound \ybmggao{} the gap is $\sim 38 \meV$~\cite{paddison_continuous_2016}.  In the related pyrochlore compounds \abo{Yb}{M} (M = Ti, Sn, Ge) an even larger gap of $\sim 50-80 \meV$~\cite{yaouanc-2013-ybsno,gaudet-2015-ybtio,hallas2016musr} is observed.  As this large energy scale stems from the atomic physics of \rth{Yb}, we expect similar crystal field energy scales in any ytterbium-based honeycomb compounds.

While the local environment is approximately cubic, the full site symmetry of the \rth{Yb} ion is lower, being only $D_{3d}$ or $C_{3v}$ (depending on the specific material considered). This lowering of symmetry splits the octahedral $\Gamma_8$ quartet into a trigonal $\Gamma_4$ doublet and a trigonal $\Gamma_5 \oplus \Gamma_6$ doublet of one-dimensional irreducible representations connected by time-reversal symmetry~\cite{huang-2014-octupolar}. Given the approximate local cubic symmetry in all of the compounds of interest, we will assume a  well isolated $\Gamma_4$ ground doublet, $\ket{\pm}$, taking the form
\begin{equation}
\label{eq:cef}
\ket{\pm} = 
\sin{\eta}\left[\cos{\zeta} \ket{\pm 7/2} 
\pm\sin{\zeta}\ket{\pm 1/2}\right] 
+ \cos{\eta}\ket{\mp 5/2},
\end{equation}
where we have chosen the quantization axis, $\vhat{z}$, along the local three-fold symmetry axis. In the spinel and breathing pyrochlore compounds, the local three-fold axis is different from site to site, while in the triangular (such as \ybmggao{}) or honeycomb compounds, it points uniformly perpendicular to the two-dimensional plane.  This form [Eq.~(\ref{eq:cef})] encompasses both the octahedral $\Gamma_6$ and $\Gamma_7$ doublets, but they do not remain distinct when the symmetry is lowered to trigonal.  We note that the angles $(\eta,\zeta)$ are somewhat redundant; mapping $(\eta,\zeta) \rightarrow (\pi-\eta,\pi+\zeta)$ only gives a redefinition $\ket{\pm} \rightarrow -\ket{\pm}$ and thus does not change any of the physics.  We can thus restrict both $\eta$ and $\zeta$ to lie between $0$ and $\pi$ without any loss of generality.

If we consider general values of $(\eta,\zeta)$, then there are two notable limits with high symmetry corresponding to the octahedral $\Gamma_6$ and $\Gamma_7$ doublets.  The $\Gamma_6$ doublet~\cite{lea1962raising}, in the notation of Eq.~(\ref{eq:cef}), corresponds to
\begin{align}
  \label{eq:cef-oct}
\eta_{\Gamma_6} &= \cos^{-1}\left(\frac{1}{3}\sqrt{\frac{35}{6}}\right),&   
\zeta_{\Gamma_6} &= \pi-\tan^{-1}\left(\sqrt{\frac{14}{5}}\right).
\end{align}
The $\Gamma_7$ doublet~\cite{lea1962raising} corresponds to the parameters
\begin{align}
  \label{eq:cef-cube}
\eta_{\Gamma_7} &= \pi - \cos^{-1}\left(\frac{1}{3}\sqrt{\frac{1}{2}}\right), &
\zeta_{\Gamma_7} &= \tan^{-1}\left(\sqrt{\frac{10}{7}}\right).
\end{align}
Note that in the case of transition metals such as the iridates or ruthenates, the $J_{\rm eff}=1/2$ states~\cite{kim-2008-spin-orbit} transform as the $\Gamma_7$ representation~\cite{bradley2010mathematical}, not the $\Gamma_6$. For the ${}^2 F_{7/2}$ manifold, the $\Gamma_7$ doublet (i.e. the analogue of the $J_{\rm eff}=1/2$ doublet) is the ground doublet if the ligands form a \emph{cube}~\cite{lea1962raising}.  This kind of cube of ligands is approximately realized in pyrochlore compounds with structural parameter $x$ close to the ideal $x_c = 3/8$~\cite{gardner-2010-review}.  Typically, rare-earth pyrochlores have $x \sim 0.32-0.34 < x_c$~\cite{gardner-2010-review}, but have ground doublets adiabatically connected to the $\Gamma_7$ state.  We note that for the case of edge-sharing perfect \emph{cubes} (as opposed to octahedra), there are two equivalent ligand paths (as in the $90^\circ$ case), but the bond angle is the tetrahedral angle $\theta_t \equiv \cos^{-1}(-1/3) \sim 109.5^\circ$ and the orientation of the ligands relative to the local axes is slightly different. We will return to this case briefly in Sec.~\ref{sec:discussion}.

We also note that there is a another high symmetry limit (somewhat) relevant for the pyrochlore compounds, with structural parameter $x = 1/4$. This corresponds to a configuration with accidental six-fold symmetry; a hexagon of ligands, with the remaining two ligands along the three-fold symmetry axis. In a point charge calculation, this gives a pure $\ket{\pm 1/2}$ ground state. This composition is somewhat stable, as the six-fold symmetry forbids mixing these states with the others of the ${}^2 F_{7/2}$ manifold. This corresponds to the crystal field parameters $\eta_{\rm hex} = \zeta_{\rm hex} = \pi/2$ in Eq.~(\ref{eq:cef}).

Since, as discussed above, the crystal field scale is very large, we can consider only a bare projection of the microscopic ion-ion interactions into these doublets. Such a model is best formulated directly in terms of the pseudo-spins
\begin{align}
\label{eq:pseudo-spin}
  S^z_i &\equiv \frac{1}{2}\left(\ket{+}_i\bra{+}_i-\ket{-}_i\bra{-}_i\right), &
  S^{\pm}_i &\equiv \ket{\pm}_i\bra{\mp}_i.
\end{align}
where the doublets $\ket{\pm}_i$ are defined at each lattice site $i$. Under crystal symmetries, the pseudo-spin operators, $\vec{S}_i$, transform in the same way as spin-1/2 operators. They are directly related to the magnetic moment $\vec{\mu}_i$ of the \rth{Yb} ion through the two $g$-factors $g_{z}$ and $g_{\pm}$, defined as
\begin{equation}
  \label{eq:moment}
  \vec{\mu}_i \equiv -g_J \mu_B P\vec{J}_iP =  -\mu_B\left[
    g_{\pm} \left( \vhat{x}_iS^x_i +
      \vhat{y}_i S^y_i \right) +g_{z}  \vhat{z}_i S^z_i \right],
\end{equation}
where $(\vhat{x}_i,\vhat{y}_i,\vhat{z}_i)$ defines a local frame with $\vhat{z}_i$ being the three-fold symmetry axis, $\vec{J}_i$ is the total angular momentum and $P$ projects into the ground doublet. The $\vhat{y}_i$ axis is defined to be along the local two-fold axis for $D_{3d}$ or perpendicular to the mirror plane for $C_{3v}$. The explicit convention for these local axes is given in App.~\ref{app:conv}. These $g$-factors are determined by the crystal field parameters $(\eta,\zeta)$ of Eq.~(\ref{eq:cef}) via
\begin{subequations}
\label{eq:g-factors}
\begin{align}
  g_{\pm} &= g_J\left[\sqrt{7} \cos \zeta  \sin (2 \eta )-4 \sin^2\zeta \sin ^2\eta\right],\\
  g_{z} &= g_J\left[\left(3 \cos (2 \zeta )+4\right) \sin ^2\eta-5 \cos ^2\eta\right],
\end{align}  
\end{subequations}
where $g_J = 8/7$ is the Land\'e $g$-factor for \rth{Yb}. Note that there are non-trivial bounds on the $g$-factors; from Eq.~(\ref{eq:g-factors}) one can show that
\begin{subequations}
\label{eq:g-factor-constraint}
\begin{align}
  -40/7 &\leq g_z \leq +8, \\
  -32/7 &\leq g_{\pm} \leq +8/\sqrt{7}.
\end{align}  
\end{subequations}

In the octahedral limit ($\Gamma_6$ doublet) defined by Eq.~(\ref{eq:cef-oct}), these $g$-factors are equal, with $g_{\pm} = g_{z} = -8/3$, both \emph{negative}.  For the limit of a cube of ligands ($\Gamma_7$ doublet) defined by Eq.~(\ref{eq:cef-cube}), the $g$-factors are given by $-g_{\pm} = g_z = 24/7$. Note that both $g$-factors can be both made positive (separately) by a redefinition of the doublet states. Given these kind of ambiguities in defining $g$-factors, it can be useful to consider quantities invariant under transformations of the doublet basis, such as $\det{\mat{g}} = g_{\pm}^2 g_z$. This gives a clear discriminant between the two cases: $\Gamma_6$ has $ g_{\pm}^2 g_z < 0$ while the $\Gamma_7$ has $g_{\pm}^2 g_z > 0$. One can use this quantity, $\sgn{(g_{\pm}^2 g_z)} = \pm 1$, more generally to give an idea whether a general doublet is closer to the $\Gamma_6$ or to the $\Gamma_7$ doublet. As an example, for the hexagonal case mentioned above, the $g$-factors are $g_{\pm} = -32/7$ and $g_z = +8/7$, corresponding to the same class, in the sense defined above, as the cubic $\Gamma_7$ doublet.~\footnote{ Note that this limit has the \emph{maximal} value for $g_{\pm}$ (though not the maximal value for the ratio $g_{\pm}/g_{z}$) and thus has XY-like moments. This may be related to the fact that the \abo{Yb}{M} pyrochlores all have relatively strong XY single-ion anisotropy.}

We should note that the two $g$-factors do not uniquely determine the composition $(\eta,\zeta)$.  Since the angular momentum $\vec{J}_i$ is only a rank-one multipole operator, the $g$-factors are not sensitive to the phases between components of the doublet separated by more than a single unit of angular momentum. This manifests in the invariance of the $g$-factors, Eq.~(\ref{eq:g-factors}), under the transformation $(\eta,\zeta) \rightarrow (\pi-\eta,\pi-\zeta)$. This transformation changes the sign of the $\ket{\pm 7/2}$ and $\ket{\mp 5/2}$ components of the ground doublet, Eq.~(\ref{eq:cef}), but not the $\ket{\pm 1/2}$ component. This invariance \emph{does not} carry over to the two-ion exchange processes. Indeed, we will see that crystal fields with the same $g$-factors can yield \emph{entirely different interactions}, due to the higher rank multipoles that are generated by the exchange processes~\cite{iwahara-2015-exchange,rau-2015-magnitude,rau-2016-order}.

\section{Two-ion physics}
\label{sec:two-ion}
We now consider the two-ion physics of the exchange interactions.  As in the one-ion case, symmetries also strongly constrain the allowed interactions between the pseudo-spins, $\vec{S}_i$ defined in Eq.~(\ref{eq:pseudo-spin}). For all the lattices of interest, at the nearest-neighbor level, such a pseudo-spin model must have the form (due to the bond symmetries)
\begin{align}
\label{eq:model}
    H_{\rm eff} &\equiv \sum_{\avg{ij}} \trp{\vec{S}}_i \mat{J}_{ij}\vec{S}^{}_j,\\
 \trp{\vec{S}}_i \vec{J}_{ij} \vec{S}^{}_j &=  J_{zz} S^z_i S^z_j -
J_{\pm}\left(S^+_i S^-_j+S^-_i S^+_j\right)+  \nonumber \\ &
J_{\pm\pm} \left(\gamma_{ij} S^+_i S^+_j+\hc \right)+ 
J_{z\pm}\left( \zeta_{ij} \left[ S^z_i S^+_j+ S^+_i S^z_j \right]+ \hc \right), \nonumber
\end{align}
where the $\gamma_{ij}$ and $\zeta_{ij}$ are bond dependent phases. This was shown for the pyrochlore case first in Ref.~[\onlinecite{curnoe-2008}], and we adopt the notation introduced in Ref.~[\onlinecite{ross-2011-quantum}]. In each case, there are three types of bonds in the local frames,  labeled $x$, $y$ and $z$ in Fig.~\ref{fig:structures}. The relevant phases factors $\gamma_{ij}$ and $\zeta_{ij}$ for these three bond types are
\begin{align}
  \gamma^{}_{x} = -\cc{\zeta}_x &= 1, &
  \gamma^{}_{y} = -\cc{\zeta}_y &= \omega, &
  \gamma^{}_{z} = -\cc{\zeta}_z &= \omega^2,
\end{align}
where $\omega = e^{2\pi i/3}$.

Our primary goal in the present work is to estimate the four exchanges $J_{zz}$, $J_{\pm}$, $J_{\pm\pm}$ and $J_{z\pm}$ in Eq.~(\ref{eq:model}) from microscopic considerations.  However, there are several equivalent ways to present the anisotropic exchange model of Eq.~(\ref{eq:model}), with each presentation offering different insights into the basic features of the model. Further, there are several dualities that map between different exchange parameter sets that are more physically transparent in one formulation over another. We thus next catalog these different representations, unifying the parametrizations used in both the quantum spin ice~\cite{gingras-2014-quantum} and Kitaev spin liquid contexts~\cite{rau-2016-arcmp}.

\subsection{Local axes}
\label{sec:non-colinear}
\begin{figure}[tp]
  \centering
  \begin{overpic}[width=0.66\columnwidth]{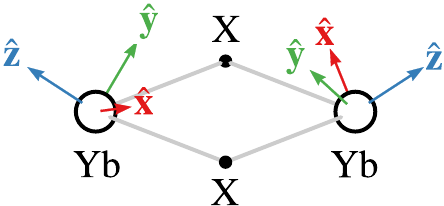}
    \put(-7,0){{\large (a)}}
  \end{overpic}
  \\  \vspace{1cm}
  \begin{overpic}[width=0.69\columnwidth]{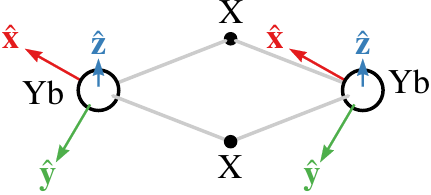}
    \put(-5,0){{\large (b)}}
  \end{overpic}\vspace{0.5cm}
  \caption{\label{fig:bond-geometry}    
    Illustration of the Yb-X-X-Yb bond geometry and local environments. In this ideal
    case the angle, $\theta$, along each Yb-X-Yb path is $90$ degrees.
    We have indicated the (a) local frames for each Yb site relevant for the pyrochlore lattices and (b) the common frames relevant for the triangular and honeycomb lattices.
  }
\end{figure}

We first consider the case where the high-symmetry axes vary from site to site. Specifically, we consider the three-fold axes, $\vhat{z}_i$, on nearest-neighbor sites to be at an angle of $\theta_{\rm t} \sim 109.45^\circ$, as is relevant to both the pyrochlore and breathing pyrochlore lattices shown in Fig. \ref{fig:bond-geometry}.  There are two alternative parametrizations of this model that will be useful to us. The first is the \emph{global} basis, where we undo the effects of the different local frames to define an overall quantization axis. We denote these global effective spins as $\bar{\vec{S}}_i$, given in terms of the local pseudo-spins as
\begin{equation}
  \bar{\vec{S}}_i \equiv \vhat{x}^{}_i S^x_i + \vhat{y}^{}_i S^y_i + \vhat{z}^{}_i S^z_i,
\end{equation}
where $(\vhat{x}_i,\vhat{y}_i,\vhat{z}_i)$ is the local frame at site $i$ (see App.~\ref{app:conv} for our conventions). Note that, due to the $g$-factors, these global pseudo-spins, $\bar{\vec{S}}_i$, are \emph{not} simply the magnetic moments $\vec{\mu}_i$ due to the non-trivial $g$-factors. Translated into this basis the symmetry allowed model, Eq.~(\ref{eq:model}), becomes
\begin{equation}
  \label{eq:global-model}
  H = \sum_{\avg{ij}} \trp{\bar{\vec{S}}}_i \bar{\mat{J}}_{ij} \bar{\vec{S}}_j,
\end{equation}
where the global exchange matrices $\bar{\mat{J}}_{ij}$ are defined as
\begin{align}
    \bar{\mat{J}}_{12} &= \left(
        \begin{array}{ccc}
             J+K &+\frac{D}{\sqrt{2}} &+\frac{D}{\sqrt{2}}  \\
             -\frac{D}{\sqrt{2}} & J & \Gamma \\
             -\frac{D}{\sqrt{2}} & \Gamma & J
        \end{array}
    \right), & 
    \bar{\mat{J}}_{13} &= \left(
        \begin{array}{ccc}
             J & -\frac{D}{\sqrt{2}} & \Gamma  \\
            +\frac{D}{\sqrt{2}} & J+K &+\frac{D}{\sqrt{2}} \\
             \Gamma & -\frac{D}{\sqrt{2}} & J
        \end{array}
    \right), \nonumber \\    
    \bar{\mat{J}}_{14} &= \left(
        \begin{array}{ccc}
             J & \Gamma & -\frac{D}{\sqrt{2}}  \\
             \Gamma & J & -\frac{D}{\sqrt{2}} \\
            +\frac{D}{\sqrt{2}} &+\frac{D}{\sqrt{2}} & J+K
        \end{array}
    \right), & 
    \bar{\mat{J}}_{23} &= \left(
        \begin{array}{ccc}
             J & -\Gamma &+\frac{D}{\sqrt{2}}  \\
             -\Gamma & J & -\frac{D}{\sqrt{2}} \\
             -\frac{D}{\sqrt{2}} &+\frac{D}{\sqrt{2}} & J+K
        \end{array}
    \right), \nonumber \\    
    \bar{\mat{J}}_{24} &= \left(
        \begin{array}{ccc}
             J &+\frac{D}{\sqrt{2}} & -\Gamma  \\
             -\frac{D}{\sqrt{2}} & J+K &+\frac{D}{\sqrt{2}} \\
             -\Gamma & -\frac{D}{\sqrt{2}} & J
        \end{array}
    \right), & 
    \bar{\mat{J}}_{34} &= \left(
        \begin{array}{ccc}
             J+K & -\frac{D}{\sqrt{2}} &+\frac{D}{\sqrt{2}}  \\
            +\frac{D}{\sqrt{2}} & J & -\Gamma \\
             -\frac{D}{\sqrt{2}} & -\Gamma & J
        \end{array}
    \right),\nonumber
\end{align}
where $\bar{\mat{J}}_{ab}$ denotes the exchange matrix between sites with sublattices $a$ and $b$.  The local parametrization of Eq.~(\ref{eq:model}) and this global parametrization are related as~\cite{ross-2011-quantum}
\begin{align}
J &= \frac{1}{3} \left(+4 J_{\pm}+2 J_{\pm\pm}+2 \sqrt{2} J_{z\pm}-J_{zz}\right), \nonumber \\
K &= \frac{2}{3} \left(-4 J_{\pm}+ J_{\pm\pm}+ \sqrt{2} J_{z\pm}+ J_{zz}\right), \nonumber \\
\Gamma &= \frac{1}{3} \left(-2 J_{\pm}-4 J_{\pm\pm}+2 \sqrt{2} J_{z\pm}-J_{zz}\right), \nonumber \\
D &= \frac{\sqrt{2}}{3} \left(-2 J_{\pm}+2 J_{\pm\pm}-\sqrt{2} J_{z\pm}-J_{zz}\right).
\end{align} 
Here, we have used a parametrization in terms of Heisenberg exchange $J$, Kitaev interaction $K$, symmetric off-diagonal exchange
$\Gamma$ and \ac{DM} interaction $D$~\footnote{This parametrization is closely related to the $J_1$, $J_2$, $J_3$ and $J_4$ introduced in Ref.~[\onlinecite{ross-2011-quantum}]. Explicitly one has $J_1=J$, $J_2=J+K$, $J_3=\Gamma$ and $J_4 = D/\sqrt{2}$. This parametrization has also been discussed briefly in Ref.~[\onlinecite{thompson2017quasiparticle}]}. There is also a non-trivial \emph{duality} in this parametrization. One notes that performing a rotation by $\pi$ about $\vhat{z}_i$ maps $S^{\pm}_i \rightarrow -S^{\pm}_i$ and $S^z_i \rightarrow +S^z_i$, we map the local exchange parameters as $(J_{zz},J_{\pm},J_{\pm\pm},J_{z\pm}) \rightarrow (J_{zz},J_{\pm},J_{\pm\pm},-J_{z\pm})$. In the global basis, this strongly mixes the four exchange constants; after such a transformation we have a new dual set of global exchanges
\begin{align}
\label{eq:pyro:dual}
\tilde{J} &= \frac{1}{9}\left(J -4K -4\Gamma + 2\sqrt{2}D\right), \nonumber \\
\tilde{K} &= \frac{1}{9}\left(-8J +5K -4\Gamma +2\sqrt{2}D\right),\nonumber \\
\tilde{\Gamma} &= \frac{1}{9}\left(-8J-4K+5\Gamma+2\sqrt{2}D\right),\nonumber \\
\tilde{D} &= \frac{1}{9}\left(2\sqrt{2}\left[2J + K +\Gamma\right] + 7D\right).
\end{align}
We thus see that there are non-trivial dual realizations of the various limits, Heisenberg, Kitaev, and so forth that are hidden in the original global representation of Eq.~(\ref{eq:global-model}). 
For example, the point $(J,K,\Gamma,D) = (-1,-8,-8,4\sqrt{2})$ maps to a dual Heisenberg antiferromagnet, with $(\tilde{J},\tilde{K},\tilde{\Gamma},\tilde{D}) \propto (1,0,0,0)$. If we further transform the $g$-factors as $(\tilde{g}_{z},\tilde{g}_{\pm}) = (g_{z},-g_{\pm})$, then the moment defined in Eq.~(\ref{eq:moment}) remains invariant. Since essentially all probes of the low-energy physics in these compounds are through some coupling to the moment $\vec{\mu}$, we see that the dual theory defined by Eq.~(\ref{eq:pyro:dual}) can be regarded as \emph{physically} equivalent to the original for most practical purposes. One can thus usually only determine the relative sign of $g_{\pm}$ and $J_{z\pm}$ from measurements at low energy (not probing the high energy crystal field levels).

We note that there is also a generalized Klein duality~\cite{kimchi-2014-3d} that is relevant in the limit of only Heisenberg and Kitaev exchange interactions~\cite{chaloupka-2010-kitaev}. By combining this with the above dualities, one can expose more Heisenberg ferro- and antiferromagnetic limits~\cite{chaloupka-2015-hidden}. We will not pursue this here, except to note that the Klein duals of the global Heisenberg ferro- and antiferromagnets are simply the \emph{local} Heisenberg ferro- and antiferromagnets with $J_{zz}=-2J_{\pm}$ and $J_{\pm\pm}=J_{z\pm}=0$.

\subsection{Uniform axes}
\label{sec:colinear}
The case where there is a uniform, global, three-fold axis is relevant to two-dimensional structures such as honeycomb or triangular lattices built from such edge-sharing octahedra (see Fig.~\ref{fig:structures}). Here, the frames can be chosen to be the same from site to site, as the three-fold symmetry axis is perpendicular to the two-dimensional plane. This basis for the exchange parameters [Eq.~(\ref{eq:model})] has so far not been used extensively~\cite{rau2014trigonal} in the literature on honeycomb or triangular Kitaev materials~\cite{rau-2016-arcmp}. However, a local basis very similar (but not identical) to that of Eq.~(\ref{eq:model}) has been used to describe \ybmggao{}~\cite{li-2015-rare-earth,li-2016-anisotropic}.  In the more commonly used basis~\cite{rau2014generic}, one has the three exchange matrices
\begin{align}
  \mat{J}_x &\equiv\left(
  \begin{array}{ccc}
    J+K& \Gamma' & \Gamma' \\
    \Gamma' & J & \Gamma \\
    \Gamma' &\Gamma & J
  \end{array}
  \right),  \nonumber \\
  \mat{J}_y &\equiv\left(
  \begin{array}{ccc}
    J & \Gamma' & \Gamma \\
    \Gamma' & J+K & \Gamma' \\
    \Gamma &\Gamma' & J
  \end{array}
  \right), \nonumber \\
  \mat{J}_z &\equiv\left(
  \begin{array}{ccc}
    J & \Gamma & \Gamma' \\
    \Gamma & J & \Gamma' \\
    \Gamma' &\Gamma' & J+K
  \end{array}
  \label{eq:colinear:conv-exch}
  \right).  
\end{align}
Note that there are no \ac{DM} interactions for the triangular and honeycomb cases due to the inversion symmetry about the bond centers. This symmetry is present in the full crystal structures of the materials of interest (we will return to the role of structural disorder in \ybmggao{} in Sec.~\ref{sec:triangular}). The more commonly used parameters are related to the exchanges defined in Eq.~(\ref{eq:model}) by
\begin{align}
J &= \frac{1}{3}\left(J_{zz}-4J_{\pm} -2J_{\pm\pm}-2\sqrt{2}J_{z\pm}\right), \nonumber\\
K &= 2 \left(J_{\pm\pm} + \sqrt{2}J_{z\pm}\right), \nonumber\\
\Gamma &= \frac{1}{3}\left(J_{zz}+2J_{\pm}+4J_{\pm\pm}-2\sqrt{2}J_{z\pm}\right), \nonumber\\
\Gamma' &= \frac{1}{3}\left(J_{zz}+2J_{\pm}-2J_{\pm\pm}+\sqrt{2} J_{z\pm}\right). 
\end{align}
As in the case with local frames discussed in Sec.~\ref{sec:non-colinear} one can obtain a duality by rotating about three-fold axis by $\pi$. This was first introduced in Ref.~[\onlinecite{chaloupka-2015-hidden}] in the context of the honeycomb iridates. This maps the exchanges $(J,K,\Gamma,\Gamma')$ to the dual exchanges
\begin{align}
\tilde{J} &= J+\frac{1}{9}\left(4K -4\Gamma+4\Gamma'\right), \nonumber \\
\tilde{K} &= \frac{1}{3}\left(-K+4\Gamma-4\Gamma'\right),\nonumber \\
\tilde{\Gamma} &= \frac{1}{9}\left(4K +5\Gamma+4\Gamma'\right),\nonumber \\
\tilde{\Gamma}' &= \frac{1}{9}\left(-2K +2\Gamma+7\Gamma'\right).
\end{align}
Note that, since the frames are the same from site to site, the pure Heisenberg limit is unaffected by this transformation (in contrast to the case with local frames).  As before, this duality exposes a number of simple hidden regimes that are not manifest in the original parametrization~\cite{chaloupka-2015-hidden}. For example, the dual of the pure Kitaev limit presents itself as combination of $J$, $K$, $\Gamma$ of $\Gamma'$ of nearly equal magnitude. As in the case of local frames, we will not explore the implications of the Klein dualities that exist for these lattices in the Heisenberg-Kitaev limit~\cite{chaloupka-2015-hidden}.

\section{Super-exchange}
\label{sec:super-exchange}
Through the results of Sec.~\ref{sec:single-ion} and Sec.~\ref{sec:two-ion}, we have outlined the generic one- and
two-ion physics of these materials.  Our goal now is to present a microscopic theoretical framework for computing the two-ion anisotropic exchange interactions, given knowledge of the single-ion crystal field ground state defined in Eq.~(\ref{eq:cef}).

To this end, we consider a pair of rare-earth ions, which we denote as $1$ and $2$, and two bridging ligands which we denote as $A$ and $B$. This exchange geometry is illustrated in Fig. \ref{fig:bond-geometry}. We will assume that super-exchange processes are driven by pathways that proceed between the rare-earth ions through the ligands. We are thus ignoring processes that involve any direct exchange between the rare-earth $4f$ orbitals (assumed to be small) or through other rare-earth orbitals, such as the $5d$ or $6s$,  of the Yb ions themselves or their associated bands in solid  (assumed to be high in energy). While the calculation for a single ligand has been described in other works~\cite{onoda-jpcm,onoda-2011-quantum,rau-2015-magnitude,rau-2016-order}, the two ligand geometry introduces new complications that deserve some attention.  

We write the Hamiltonian of this system as
\begin{equation}
  H_0 \equiv H_{f,1}+H_{f,2} + H_{p,A}+ H_{p,B},
\end{equation}
where $H_{f,1}$ and $H_{f,2}$ are atomic Hamiltonians for each of the two rare-earth ions while $H_{p,A}$ and $H_{p,B}$ are for the two ligand sites. On the two ligand sites, we consider only the cost of a single hole on a ligand (the atomic potential), defined as $\DeltaOx$,  and the (additional) cost to place two holes together on the same ligand, which we define as $U_p$.  The rare-earth atomic physics of $H_{f,1}$ and $H_{f,2}$ is discussed in some detail in App.~\ref{app:atomic}.  We will not invoke the form of the crystal field part of the rare-earth atomic Hamiltonian aside from the fact that its ground doublet is approximately as given in Eq.~(\ref{eq:cef}), ignoring its effects on the virtual states involved in the super-exchange (the plausibility of the approximation is discussed in App.~\ref{app:atomic}).  We perturb the atomic Hamiltonian $H_0$ with the hybridization terms
\begin{equation}
  \label{eq:hyb}
V \equiv \sum_{\alpha\beta} \sum_{\lambda=A,B} \left[
 {t}_{1\lambda}^{\alpha\beta} \h{f}_{1\alpha} p^{}_{\lambda,\beta} +
 {t}_{2\lambda}^{\alpha\beta} \h{f}_{2\alpha} p^{}_{\lambda,\beta} + \hc
\right],
\end{equation}
that represent electron hopping between the orbitals of the rare-earth and ligand ions. The hopping matrices $t_{1\lambda}$ and $t_{2\lambda}$ can be estimated via a Slater-Koster type approach \cite{slater-1954-simplified}.  They depend on the local frames at site $1$ and site $2$ as well as the overlap parameters $t_{pf\sigma}$ and $t_{pf\pi}$. Generically, we expect that $|t_{pf\sigma}| > |t_{pf\pi}|$ and that they have opposite sign. A typical ratio that we will use is  $t_{pf\pi}/t_{pf\sigma} \sim -0.3$, though most of our results are not particularly sensitive to its precise value.  Note that both the $\sigma$ and $\pi$ overlap are involved even in the ideal geometry with a $90^{\circ}$ X-Yb-X bond angle. While one cannot construct an exchange path using only $t_{pf\sigma}$ (in the ideal case), combinations of $t_{pf\pi}$ and $t_{pf\sigma}$ do contribute. This is a \emph{key difference} from the case considered in Ref.~[\onlinecite{jackeli-2009-mott}] for transition metal oxides where the restriction to the $T_{2g}$ states of the $d$ manifold allows only a single hopping parameter to appear. In the calculations detailed below, the $t_{1\lambda}$ and $t_{2\lambda}$ matrices do not appear independently, but only in the combinations
\begin{equation}
  T_{\lambda} \equiv t^{}_{1\lambda} \h{t}_{2\lambda},
\end{equation}
where $\lambda=A,B$. Note that these matrices are symmetric, $\trp{T}_{\lambda} = T^{}_{\lambda}$, for all cases of interest.

\begin{figure}[tp]
\includegraphics[width=0.8\columnwidth]{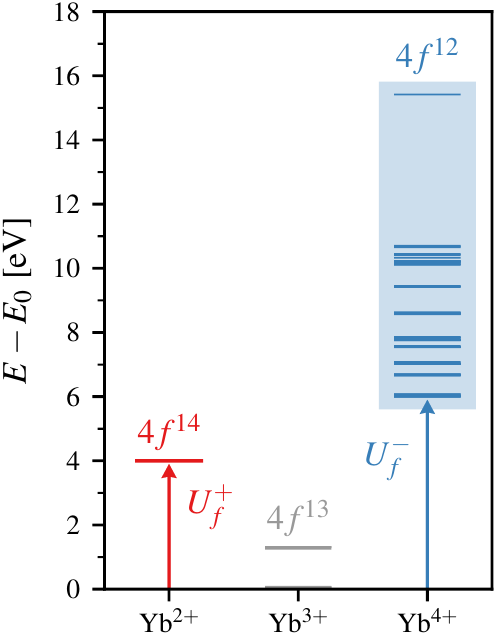}
\hspace{1cm}
  \caption{\label{fig:levels}    
    Schematic energy levels of ytterbium ions in solid, relative to the \rth{Yb} ion ground state energy $E_0$, including the Yb\tsup{2+}, Yb\tsup{3+} and Yb\tsup{4+} valences. The levels of Yb\tsup{4+} are shown for the atomic parameters described in App.~\ref{app:atomic} with a trigonal (point-charge) crystal field added to illustrate the scale of these splittings. Minimal charge transfer energies $U^{\pm}_f \equiv E_0(4f^{13 \pm 1})-E_0$ are indicated.
  }
\end{figure}

In this approach, super-exchange interactions are generated at fourth-order in perturbation theory in the ligand-rare-earth hybridization~\cite{onoda-2011-quantum}.  Given the complexity of the rare-earth site Hamiltonians $H_{f,1}$ and $H_{f,2}$, performing the fourth order perturbation theory is  analytically challenging. To proceed, we will first notice that the hybridization perturbation $V$ necessarily changes the charge state of the \rth{Yb} ion, connecting the $\f^{13}$ manifold to the $\f^{14}$ or $\f^{12}$ manifolds. Since the $\f^{14}$ manifold is simply a closed shell, it produces particularly simple contributions in perturbation theory. We define the energy cost to excite from the $\f^{13}$ ground state to the $\f^{14}$ state as $U^+_f$. The $\f^{12}$ manifold has some internal structure, with a total of ${14 \choose 2} = 91$ states. Keeping only the free-ion interactions, ignoring any crystal field splittings, these are distributed among $13$ distinct energy levels~\cite{meftah-2013-spectrum}.  The composition and position of these levels is set by the atomic physics of Yb\tsup{4+}, namely through the Coulomb interaction encoded in the Slater integrals $F_2$, $F_4$ and $F_6$ as well as in the spin-orbit coupling $\zeta_{\rm SO}$. We denote the \emph{minimal} excitation energy from the $\f^{13}$ ground state to the $\f^{12}$ manifold as $U^-_f$; the full spectrum will then have the form $U^-_f + E_a$ where the $E_a$ are the energies of the Yb\tsup{4+} ion ($4f^{12}$). The required single-ion energies and states of the $\f^{12}$ configuration can be computed using diagonalization in this $91$-dimensional subspace with an appropriate choice of atomic parameters (see App.~\ref{app:atomic} for details)~\cite{meftah-2013-spectrum}. The free-ion energy level scheme for Yb$^{2+}$, Yb$^{3+}$ and Yb$^{4+}$ is illustrated in Fig.~\ref{fig:levels}.

In contrast to the cases considered in Refs.~[\onlinecite{onoda-2011-quantum},\onlinecite{rau-2015-magnitude}] the presence of two equally spaced ligands bridging the rare-earth ions leads to additional exchange pathways. Explicitly, we consider the fourth-order processes defined by the operator~\cite{lindgren-1974-rayleigh}
\begin{equation}
  H_{\rm eff} = PVRVRVRVP,
\end{equation}
where $P$ projects into the low-energy subspace of crystal field ground doublets at each site, $R$ is the resolvent of the rare-earth and ligand atomic states and $V$ is the perturbing hybridization given in Eq.~(\ref{eq:hyb}). For virtual states involving only $f^{14}$ configurations, the resolvent $R$ is trivial and (effectively) proportional to the identity.  The processes that involve $f^{12}$ virtual states always have the ligands in their ground state with one rare-earth ion in an $f^{12}$ configuration and the other in a $f^{14}$ configuration.  The corresponding resolvent is then
\begin{equation}
   R = \sum_{a \in f^{12}} \frac{P^-_{1,a} P^+_2 + P^+_1 P^{-}_{2,a}}{U_f^+ +U_f^- + E_a} \equiv
   \frac{Q_1 + Q_2}{U_f^+ + U_f^-},
\end{equation}
where the sum runs over the distinct energy levels $U^-_f+E_a$ of the $f^{12}$ configuration (relative to the $f^{13}$ ground state) and $P^-_a$ projects into the subspace of the $E_a$ level. The operator $P^+$ projects into the closed-shell $f^{14}$ state.  We have factored out a $1/(U_f^+ +U_f^-)$ to define dimensionless resolvents $Q_1$, $Q_2$ for each site.  Taking $E_a=0$ recovers the so-called charging approximation used in Refs.~[\onlinecite{onoda-2011-quantum}, \onlinecite{rau-2015-magnitude}] since $\sum_{a \in f^{12}} P^-_a$ simply projects into the $f^{12}$ manifold.

The types of processes that involve only a single ligand have been discussed in Refs.~[\onlinecite{onoda-2011-quantum}, \onlinecite{rau-2015-magnitude}]. We go through the details of all the processes involved in the two ligand case in App.~\ref{app:processes}.  The final effective Hamiltonian for the pair of sites takes the form
\begin{equation}
\label{eq:heff}
H_{\rm eff} =
 \sum_{\alpha\beta\mu\nu}
\left[
 \mathcal{I}^{\alpha\beta\mu\nu}  O^{\alpha\beta}_1 O^{\mu\nu}_2
+\mathcal{K}^{\alpha\beta\mu\nu}\left( {O}^{\alpha\beta}_1 \tilde{O}^{\mu\nu}_2+
\tilde{O}^{\alpha\beta}_1 {O}^{\mu\nu}_2\right)
\right],
\end{equation}
where at each site we have defined the operators
\begin{subequations}
\label{eq:o-defn}
\begin{align}
  O^{\alpha\beta} &\equiv P \h{f}_\alpha f^{}_\beta P, \\
  \tilde{O}^{\alpha\beta} &\equiv P \h{f}_\alpha Q f^{}_\beta P.
\end{align}
\end{subequations}
The super-exchange tensors $\mathcal{I}$ and $\mathcal{K}$ have the form
\begin{subequations}
  \label{eq:exchange-tensors}
\begin{align}
  \mathcal{I}^{\alpha\beta\mu\nu} &\equiv 
2
\sum_{\lambda=A,B} \left[
\frac{
\left(
1-\kappa
\right)  T_{\lambda}^{\alpha\nu}
  \left[\h{T}_{\lambda}\right]^{\mu\beta} +
   T_{\lambda}^{\alpha\nu} \left[\h{T}_{\cb{\lambda}}\right]^{\mu\beta}
}{(U^+_f +\DeltaOx)^3}
\right],\\
  \mathcal{K}^{\alpha\beta\mu\nu} &\equiv 
\sum_{\lambda=A,B} \left[
\frac{
  T_{\lambda}^{\alpha\nu}
  \left[\h{T}_{\lambda}\right]^{\mu\beta} +
   T_{\lambda}^{\alpha\nu} \left[\h{T}_{\cb{\lambda}}\right]^{\mu\beta}
}{(U^+_f+\DeltaOx)^2(U^+_f + U^-_f)}
\right],
\end{align}
\end{subequations}
where we have defined  the parameter $\kappa$ as 
\begin{equation}
  \kappa \equiv  \frac{U_p}{2(U^+_f +\DeltaOx) + U_p}. 
\end{equation}
Generally, we expect $U_p \lesssim 2(U^+_f +\DeltaOx)$ and thus $\kappa \lesssim 1$.  More suggestively, these expressions can be written in terms of the total hopping $T \equiv \sum_{\lambda = A,B} T_{\lambda}$ as
\begin{subequations}
\begin{align}
  \mathcal{I}^{\alpha\beta\mu\nu} &\equiv 
\frac{2}{(U^+_f +\DeltaOx)^3}\left(
T^{\alpha\nu} \left[\h{T}\right]^{\mu\beta}
-\kappa \sum_{\lambda=A,B} T_{\lambda}^{\alpha\nu}
  \left[\h{T}_{\lambda}\right]^{\mu\beta} \right)
,\\
  \mathcal{K}^{\alpha\beta\mu\nu} &\equiv 
\frac{T^{\alpha\nu} \left[\h{T}\right]^{\mu\beta}}
{(U^+_f+\DeltaOx)^2(U^+_f + U^-_f)}.
\end{align}
\end{subequations}
Thus, if we further take $\kappa \ll 1$, we can express the exchange entirely in terms of the total hopping $T$.  This is reminiscent of a common approximate treatment of this physics~\cite{santini-2009-review,iwahara-2015-exchange} which first integrates out the ligands to generate an effective $f$-$f$ hopping then considers super-exchange physics in this setting. In our approach this corresponds to $U^{\pm}_f \ll \DeltaOx$, keeping only  the leading terms in $1/\DeltaOx$. Effectively, this corresponds to taking $\mathcal{I} \ll \mathcal{K}$. However, given that there is not a clear separation of scales between $U^{\pm}_f$ and $\DeltaOx$, we will consider such a limit only for illustrative purposes.

Given the rarity of tetravalent ytterbium, as well as the tendency for Yb to have valence fluctuations between trivalent and divalent states in intermetallic compounds~\cite{lawrence-1981-valence}, there is another useful artificial limit obtained by excluding the $\f^{12}$ states entirely. This corresponds to taking $\DeltaOx, U^+_f \ll U^-_f$ and thus having $\mathcal{K} \ll \mathcal{I}$. The exchange physics is much simpler here; the resolvent $Q$ is removed and the atomic energy scales enter only through the overall scale $(U^+_f + \DeltaOx)^{-3}$ (unimportant for determining the specific anisotropic exchange regime of interest) and the ratio $\kappa$.

Neither of these simplified limits are sufficiently realized to be used reliably. To see this, consider estimates for the various atomic parameters that appear here, such as $U^{\pm}_f$, $\Delta$, $U_p$ and the energy splittings of the $f^{12}$ intermediate states. First, note that the energies that appear in the resolvent $Q$ can safely be set to their free-ion values, given the screening of the higher Coulomb integrals, $F_2$, $F_4$ and $F_6$ is not usually significant~\cite{van-der-marel-1988-atomic}. The parameters $U^{\pm}_f$ are more difficult to obtain. Estimates from various spectroscopic probes in rare-earth \emph{metals}~\cite{van-der-marel-1988-atomic} give estimates of $U^+_f + U^-_f \sim 7\ {\rm eV}$ or so across the entire series of rare-earth ions. This value is strongly reduced from their bare (free-ion) values by screening effects. In insulating rare-earth compounds, one may expect screening to be somewhat less effective than in metals. Given the paucity of information on the $U^{\pm}_f$ parameters, we adopt the values
\begin{equation}
  U^+_f = U^-_f \sim 5\ {\rm eV},
\end{equation}
This gives $U^+_f + U^-_f \sim 10\ {\rm eV}$, somewhat higher than the value for metals found in Ref.~[\onlinecite{van-der-marel-1988-atomic}]. We choose $U^+_f = U^-_f$ purely for convenience; noting that it is at odds with the expectation of $U^+_f < U^-_f$ from chemical disfavoring of tetravalent Yb in materials. We estimate the ligand parameters from related oxides; ab-initio band-structure calculations for the series \abo{R}{Ti}~\cite{xiao2007theoretical,deilynazar2015first} gives a gap between the rare-earth $f$ states and ligand $p$ states as being $\Delta \sim 4\ {\rm eV}$.  For simplicity, we will assume this remains true for the various (non-oxygen) ligands we consider below. For the repulsion, we use a value $U_p \sim 3\ {\rm eV}$ for both oxide and chalcogenide cases.

This set of numbers clearly shows that all the different processes described in the previous section can appear on equal footing with respect to the basic energy scales involved. Unless otherwise stated, we will use the above atomic parameters when computing the exchanges and comparing to materials. To obtain the exchange parameters we simply compute the super-exchange tensors $\mathcal{I}$ and $\mathcal{K}$, as defined in Eq.~(\ref{eq:exchange-tensors}) and compute the projected operators $O^{\alpha\beta}$ and $\tilde{O}^{\alpha\beta}$ for all of the $4f$ states, combining them as in Eq.~(\ref{eq:heff}). The projected operators map to the pseudo-spins [Eq.~(\ref{eq:pseudo-spin})] of the crystal field doublet
\begin{align}
    O^{\alpha\beta} &=u^{0}_{\alpha\beta} + \vec{u}_{\alpha\beta}\cdot \vec{S}, &
    \tilde{O}^{\alpha\beta} &=\tilde{u}^{0}_{\alpha\beta} + \tilde{\vec{u}}_{\alpha\beta}\cdot \vec{S},
\end{align}
where the $u$, $\tilde{u}$ parameters depend on the atomic states $\alpha,\beta$ as well as the composition of the crystal field doublet encoded in $(\eta,\zeta)$. The constant pieces can be discarded and the remaining factors then give an anisotropic exchange model, as described in Sec.~\ref{sec:two-ion}, from which we can extract the symmetry allowed exchange parameters.

\section{Validation}
\label{sec:validation}
\begin{figure}[tp]
  \centering
  \includegraphics[width=0.8\columnwidth]{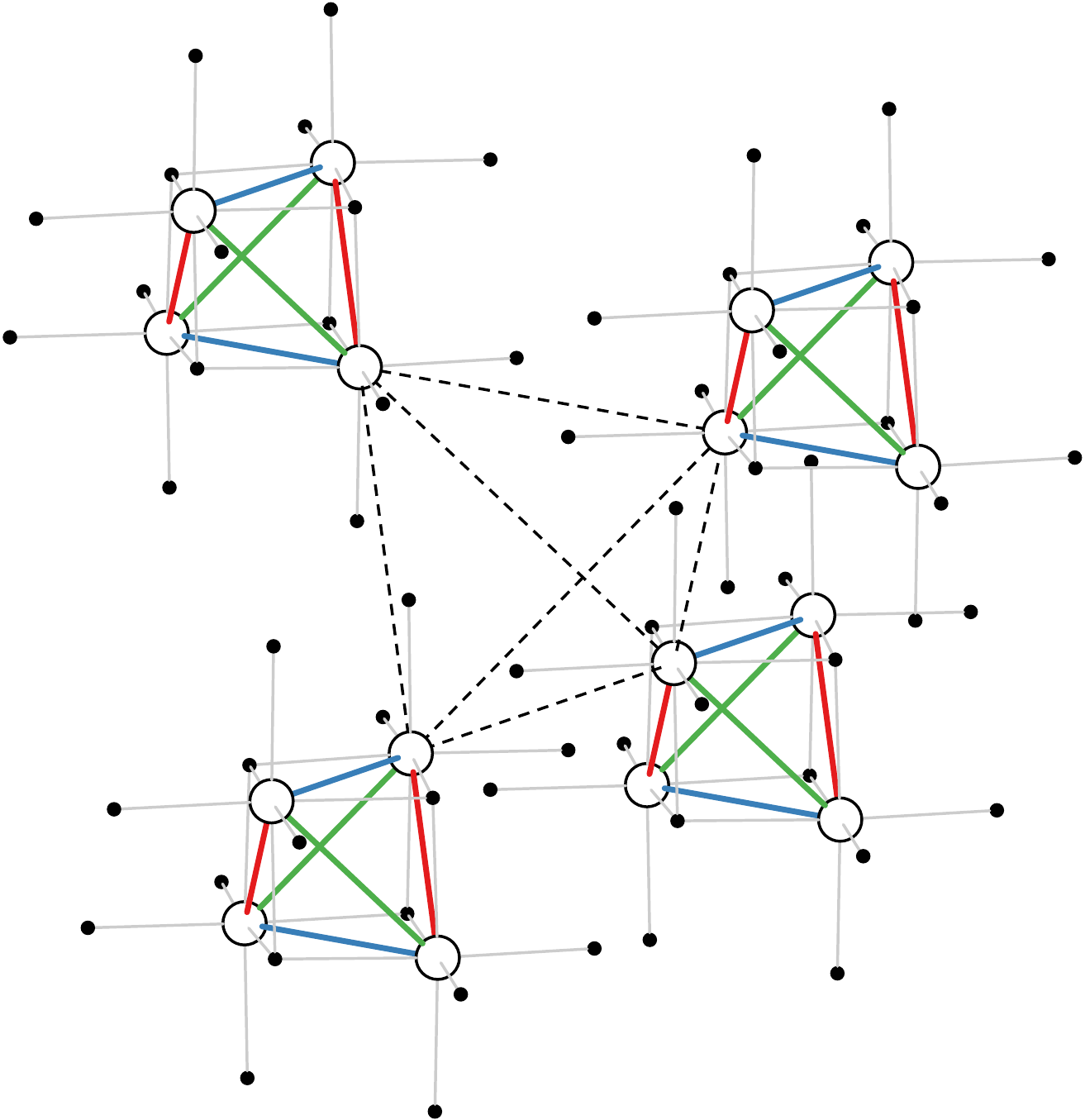}
  \caption{\label{fig:structure:breathing-pyrochlore}    
    Crystal structure of breathing pyrochlore lattice as in \byzo{}. We show the lattice of nearly independent tetrahedra formed by rare-earth ions (open circles) and the ligands which sit at the corners of the cube defined by these tetrahedra. In the ideal case the ligands form a perfect octahedron around each rare-earth ion and have a bond angle of $90^\circ$ . The dashed lines show the large tetrahedra that connect the smaller tetrahedral units.
  }
\end{figure}

We now validate this theoretical methodology in detail for the breathing pyrochlore compound \byzo{}. From the results of comparisons to experimental data presented in Refs.~[\onlinecite{rau-2016-byzo,haku-2016-byzo,park-2016-byzo}], one finds a dominant antiferromagnetic Heisenberg interaction $J$ and large (indirect) DM interaction $D$; specifically, one finds~\cite{rau-2016-byzo}
\begin{align}
  \label{eq:best-fit-global}
  J      &\sim +0.592 \meV, &
  K      &\sim -0.011\meV, \nonumber \\
  \Gamma &\sim -0.010 \meV, &
  D      &\sim -0.164 \meV.
\end{align}
The results of Refs.~[\onlinecite{haku-2016-byzo}, \onlinecite{park-2016-byzo}] are qualitatively (and essentially quantitatively) identical. In the local language, this corresponds to
\begin{align}
  J_{zz} &= -0.040 \meV, & 
  J_{\pm} &= +0.140 \meV, & \nonumber \\
  J_{\pm \pm} &= +0.160 \meV,& 
  J_{z\pm} &=  +0.302 \meV.
\end{align}
For the $g$-factors, after further refinement using data in a magnetic field, one finds a (weak) Ising anisotropy, with $(g_z,g_{\pm}) = (2.72,2.30)$~\cite{rau-2018-byzo} where we have left the signs arbitrary. For a given pair of $g$-factors, there are only a handful of consistent crystal field composition parameters~\cite{rau-2016-byzo}. We can thus hope to find what values of $(\eta,\zeta)$ to use for \byzo{} from the $g$-factors. There is some ambiguity here; first, there is the duality discussed in Sec.~\ref{sec:single-ion} and Sec.~\ref{sec:two-ion} that maps $g_{\pm} \rightarrow - g_{\pm}$ and $J_{z\pm} \rightarrow -J_{z\pm}$. Second, we can map $(g_z,g_{\pm}) \rightarrow (-g_z, -g_{\pm})$ without affecting the low energy physics. The determination of the crystal field composition parameters (and thus the exchange parameters) \emph{is}, however, sensitive to these signs. With these redundancies in mind, there are a total of eight different possible crystal field compositions that are consistent with the experimentally determined $g$-factors of
\byzo{}.

To further narrow down the possible crystal field compositions, we will only consider those which give signs for the $g$-factors that match that of the $\Gamma_6$ doublet expected in the ideal octahedral limit. That is, we only consider solutions where both $g_z$ and $g_{\pm}$ are \emph{negative}. In particular, this expectation can be corroborated through a point charge calculation of the crystal field Hamiltonian using the local ligand geometry~\cite{kimura-2014-byzo} of \byzo{}. Such a calculation does not produce a quantitatively correct level structure, finding excitations at $25\meV$, $31\meV$ and $70\meV$ instead of the $38\meV$, $54\meV$ and $68\meV$ seen experimentally. It does, however, produce a ground doublet with $g_z \sim -2.59 < 0$ and $g_{\pm} \sim -2.70 < 0$.  It thus appears reasonable to expect that the correct crystal field composition parameters share these signs for the $g$-factors. This narrows the possible doublet compositions consistent with the $g$-factors to just two
\begin{subequations}
\begin{align}
  (\eta_1,\zeta_1) &= (+0.716,+1.692), \label{eq:byzo-g-one} \\
  (\eta_2,\zeta_2) &= (+2.426,+1.450).\label{eq:byzo-g-two}
\end{align}
\end{subequations}
For both of these crystal field compositions, we can compute the expected exchange constants within the framework of Sec.~\ref{sec:super-exchange}. Since we are comparing directly with \byzo{}, we use the true bond angle of $92.94^\circ$ rather than the idealized $90^\circ$.  For the Slater-Koster ratio we use $t_{pf\pi}/t_{pf\sigma} = -0.3$, and the atomic parameters given in Sec.~\ref{sec:super-exchange}. We find that the first solution [Eq.~(\ref{eq:byzo-g-one})] gives a dominant antiferromagnetic Heisenberg interaction with a sub-dominant DM interaction and small symmetric anisotropies. The second solution gives a dominant \ac{DM} interaction with the remaining sub-dominant exchanges being roughly equal.  Note that this is a striking example of two systems with identical $g$-factors, but wildly different anisotropic exchanges.

\begin{figure}[tp]
  \centering
  \includegraphics[width=0.9\columnwidth]{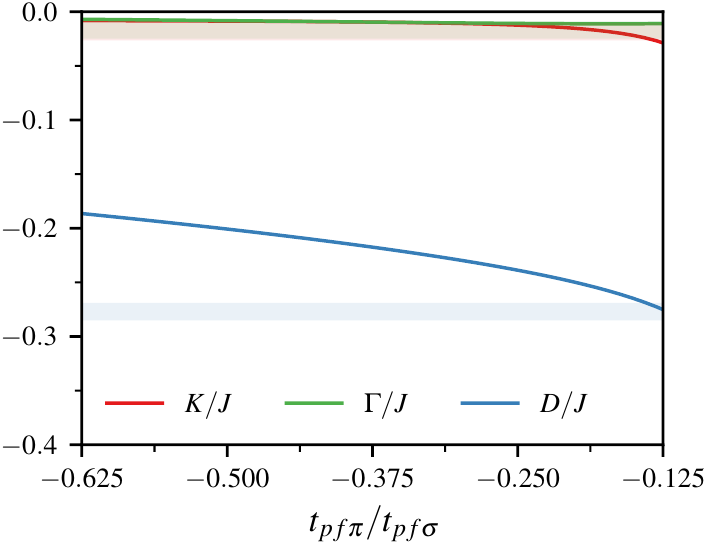}
  \caption{\label{fig:byzo-sk-ratio}
    Variation of the exchange constants for \byzo{} as a function of the Slater-Koster overlap ratio $t_{pf\pi}/t_{pf\sigma}$. For all values showing the Heisenberg interaction, $J$, is dominant and antiferromagnetic. The leading sub-dominant part is an (indirect) \ac{DM} interaction ($D < 0$) of magnitude $|D|/J \sim 0.2-0.3$. Over the entire range the symmetric anisotropies, the Kitaev interaction $K$ and off-diagonal term $\Gamma$ are negligible relative to $J$ and $D$. We have shaded the experimentally determined~\cite{rau-2016-byzo} ratios given by Eq.~(\ref{eq:best-fit-global}).
  }
\end{figure}

We thus assign \byzo{} to the region near $(\eta,\zeta) = (+0.716,+1.692)$ [Eq.~(\ref{eq:byzo-g-one})], given that it has the same sign structure for its $g$-factors as the ideal cubic limit and produces exchange in semi-quantitative agreement with the values fitted from experiment~\cite{rau-2016-byzo,haku-2016-byzo,park-2016-byzo}. Explicitly one finds that $J>0$ and
\begin{align}
  K/J &= -0.014, & \Gamma/J &= -0.011, & D/J &= -0.228.
\end{align}
The exchanges obtained here are fairly \emph{insensitive} to the detailed parameter choices made in the calculation. As an example, consider the variation of these exchanges with the Slater-Koster ratio, $t_{pf\pi}/t_{pf\sigma}$, shown in Fig.~\ref{fig:byzo-sk-ratio}. Over the entire range, one finds that the regime with large antiferromagnetic exchange and subdominant DM interaction is maintained.
We have checked that this remains true under small variations of the crystal field compositions and the various atomic parameters $U^{\pm}_f$, $\Delta$ and $U_p$ as well. We thus see that for \byzo{}, the exchange regime is robust to changes in both the theoretical parameters as well those extracted from experiment.

Given there is some uncertainty in the $g$-values, the Slater-Koster ratio and the atomic parameters, we do not attempt to tune these numbers to reproduce the fitted exchanges. For example, Ref.~[\onlinecite{park-2016-byzo}] reports $g$-factors (ignoring the signs) of $(g_z,g_{\pm}) = (3.0,2.4)$ (from neutron scattering) and $(g_z,g_{\pm}) = (2.54,2.13)$ (from electron paramagnetic resonance), while Ref.~[\onlinecite{rau-2016-byzo}] reports $(g_z,g_{\pm})=(3.0,2.36)$ and Ref.~[\onlinecite{haku-2016-byzo}] finds $(g_z,g_{\pm}) = (2.22,2.78)$. Given each pair of $g$-factors implies a different set of possible crystal field compositions, we will be content with the fact that the regime of $J > |D| \gg K, \Gamma$ exists for crystal field compositions that are reasonable for \byzo{}.

\section{Cubic limits}
\label{sec:cubic-limit}
With the expression, Eq.~(\ref{eq:exchange-tensors}), for the exchange interactions validated for \byzo{}, we now apply this framework for the variety of crystal structures discussed in the Introduction. We first look at the cubic limits where the doublets are $\Gamma_6$ or $\Gamma_7$.  For simplicity, we work in the limit of ideal $90^\circ$ bond angle and use the atomic parameters defined in Sec.~\ref{sec:super-exchange}.  We consider arbitrary Slater-Koster ratios, using the short-hand $\rho \equiv t_{pf\pi}/t_{pf\sigma}$.  In these two limits (with an appropriate choice of ground doublet basis), the distinction between the two cases (uniform and local axes) no longer exists. We will thus discuss the results for $\Gamma_6$ in the global frame and in the appropriate dual global frame for $\Gamma_7$, where the
exchange interactions computed in each case can be directly compared. This coincidence implies that only $J$, $K$ or $\Gamma$ can be non-zero; as the $D$ and $\Gamma'$ interactions are not shared between the two different parametrizations. Alternatively, one can note that the ideal limit has higher (accidental) symmetry (inversion about the bond center and a reflection symmetry) that force $D = 0$ and $\Gamma'=0$.

For the $\Gamma_6$ doublet, we find that (with $t_{pf\sigma}$ in eV)
\begin{align}
  J(\Gamma_6) &=  t_{pf\sigma}^4\left(
                5.955\rho^2
                - 4.010 \rho^3
                + 0.554 \rho^4
                \right), \nonumber \\
  K(\Gamma_6) &=  t_{pf\sigma}^4\left(
                0.071\rho^2
               + 0.099 \rho^3
                + 0.027 \rho^4
                \right), \nonumber\\
  \Gamma(\Gamma_6) &= 0. \nonumber                 
\end{align}
We thus see that for the usual $|\rho| \lesssim 1$, the exchange is \emph{strongly} Heisenberg-like and antiferromagnetic, with $K(\Gamma_6) \ll J(\Gamma_6)$, and the symmetric off-diagonal exchange $\Gamma(\Gamma_6)$ is zero~\footnote{This absence of $\Gamma$ is also found in the transition metal case with $J_{\rm eff}=1/2$ doublets~\cite{jackeli-2009-mott,rau2014generic}}. For the $\Gamma_7$ doublet we find somewhat similar results
\begin{align}
  J(\Gamma_7) &=  t_{pf\sigma}^4\left(
                -0.012\rho^2
                +0.008 \rho^3
                +0.315 \rho^4
                \right), \nonumber \\
  K(\Gamma_7) &=  t_{pf\sigma}^4\left(
                0.036\rho^2
               + 0.038 \rho^3
                - 0.016 \rho^4
                \right), \nonumber \\
                  \Gamma(\Gamma_7) &= 0. \nonumber  
\end{align}
The overall scale for the $\Gamma_7$ doublet is several orders of magnitude smaller in absolute terms than the $\Gamma_6$ case. These exchanges are also \emph{significantly} more anisotropic, with the antiferromagnetic Kitaev interactions dominating somewhat over the antiferromagnetic Heisenberg exchange, i.e. $J(\Gamma_7)/K(\Gamma_7) = 0.51$ for $\rho = -0.3$. Explicitly, if $|\rho| \ll 1$ then
\begin{subequations}
\begin{align}
  K(\Gamma_6)/J(\Gamma_6) &\sim 1.2 \cdot 10^{-2}, \\
  K(\Gamma_7)/J(\Gamma_7) &\sim -3.07, \\
  J(\Gamma_7)/J(\Gamma_6) &\sim -2.0 \cdot 10^{-2} .
\end{align}
\end{subequations}
We thus see that the separation of scales between $J$ and $K$ and between the overall scales of $\Gamma_6$ and $\Gamma_7$ remains as $\rho \rightarrow 0$. Note that the absence of $O(\rho^0)$ terms in the polynomials indicates that $t_{pf\sigma}$ overlap alone cannot induce exchange interactions (as mentioned in Sec.~\ref{sec:super-exchange}). Further, a pair of $t_{pf\pi}$ overlaps are needed, as indicated by the lack of the $O(\rho)$ term.

We can better understand these results by considering some of the artificial limits discussed in Sec.~\ref{sec:super-exchange}. To this end, we decompose the super-exchange into two parts: one coming from the $\mathcal{I}$ parts and one coming from $\mathcal{K}$ parts in Eq.~(\ref{eq:exchange-tensors}). First, consider the $\Gamma_6$ limit for which one has
\begin{align}
  J_{\mathcal{I}}(\Gamma_6) &=  t_{pf\sigma}^4\left(
                3.397\rho^2
                - 2.080 \rho^3
                + 0.318 \rho^4
                \right), \nonumber \\
  K_{\mathcal{I}}(\Gamma_6) &= 0; \nonumber \\
  J_{\mathcal{K}}(\Gamma_6) &=  t_{pf\sigma}^4\left(
                2.558\rho^2
               -1.929 \rho^3
                + 0.235 \rho^4
                \right), \nonumber \\
  K_{\mathcal{K}}(\Gamma_6) &= t_{pf\sigma}^4\left(
                0.071\rho^2
               + 0.099 \rho^3
                + 0.027 \rho^4
                \right). \nonumber
\end{align}
We thus see that the contributions from $\mathcal{I}$ are entirely isotropic; the finite Kitaev interaction stems from the $\mathcal{K}$ parts that involve the $f^{12}$ intermediate states. One should note, however, that the Heisenberg part receives roughly equal contributions from both the $\mathcal{I}$ and $\mathcal{K}$ channels. This is not the case for the $\Gamma_7$ doublet, where one finds that
\begin{align}
  J_{\mathcal{I}}(\Gamma_7) &=   t_{pf\sigma}^4\left(0.162 \rho^4\right), \nonumber \\
  K_{\mathcal{I}}(\Gamma_7) &= 0, \nonumber \\
  J_{\mathcal{K}}(\Gamma_7) &=  t_{pf\sigma}^4\left(
                -0.012\rho^2
                +0.0082 \rho^3
                +0.152\ \rho^4
                \right), \nonumber \\
  K_{\mathcal{K}}(\Gamma_7) &= t_{pf\sigma}^4\left(
                0.036\rho^2
               + 0.038 \rho^3
                - 0.016 \rho^4
                \right). \nonumber
\end{align}
Here we see that the $\mathcal{I}$ parts are purely isotropic, as for $\Gamma_6$, while the $\mathcal{K}$ parts are now mostly Kitaev.

These results for the $\Gamma_6$ and $\Gamma_7$ limits can be understood in a way similar to the $J_{\rm eff}=1/2$ case for transition metal spin-orbit Mott insulators~\cite{rau-2016-arcmp}. Essentially, this is a reflection of the fact that this calculation is related to the analogous one when taking the opposite order of limits; taking the crystal field and spin-orbit energy scales to be large first. This leads to a single half-filled band for the crystal field ground doublet. Due to the extra symmetries enjoyed by the ideal pair of edge-shared octahedra, there is only a single, pseudo-spin independent hopping amplitude. Further, since this is effectively a single band model, there is only an effective Hubbard like on-site interaction.  Thus in this (very artificial) limit, one naturally obtains a pseudo-spin rotationally invariant antiferromagnetic Heisenberg interaction. If one were to apply this same logic for edge-shared octahedra for $J_{\rm eff}=1/2$ states~\cite{jackeli-2009-mott}, one finds a complete cancellation of the hopping amplitude~\cite{shitade-2009-kitaev} and thus \emph{no} generation of exchange. For a Yb ion, this does \emph{not} occur for a $\Gamma_6$ doublet, the ground state for an ideal octahedral cage (as seen in the explicit expressions for the exchange constants). One thus  expects to obtain a robust Heisenberg antiferromagnet.  For the $\Gamma_7$ doublet, the analogue of the $J_{\rm eff}=1/2$ doublet, one \emph{does} find a similar cancellation and the overall exchange scale is  strongly suppressed. Just as in the transition metal case, the exchange interactions are determined by subleading parts of the super-exchange~\cite{jackeli-2009-mott}. In this regime, the exchange constants are expected to be sensitive to the details of the super-exchange calculation, such as the Slater-Koster ratio, $t_{pf\pi}/t_{pf\sigma}$, and the atomic energies $U^{\pm}_f$ and $\DeltaOx$.

\begin{figure}[tp]
  \centering
  \includegraphics[width=0.9\columnwidth]{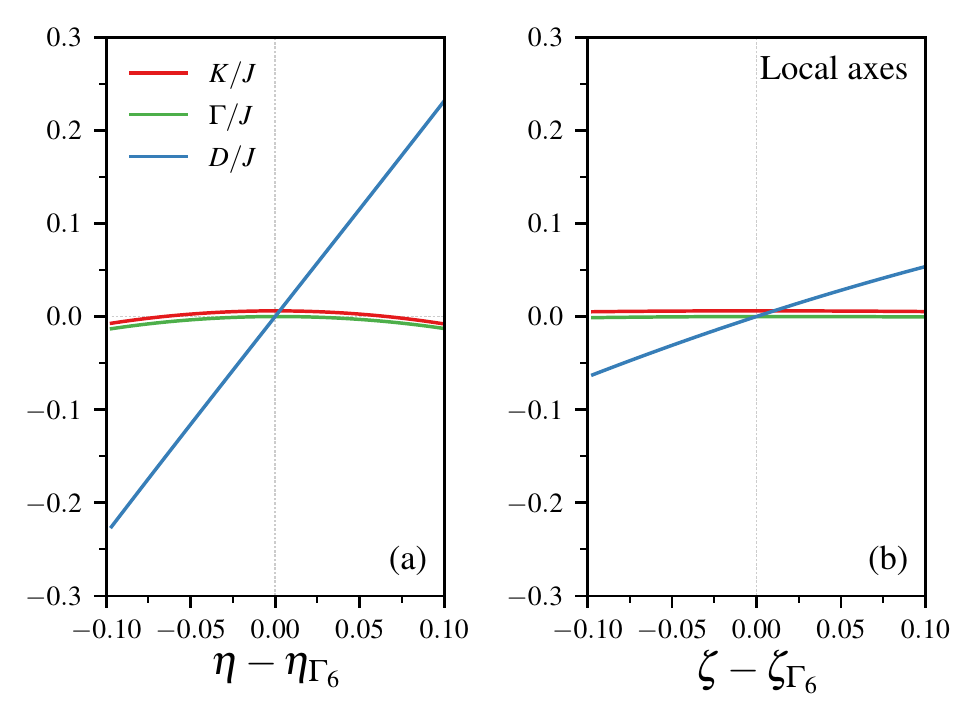}\\
  \vspace{0.25cm}
  \includegraphics[width=0.9\columnwidth]{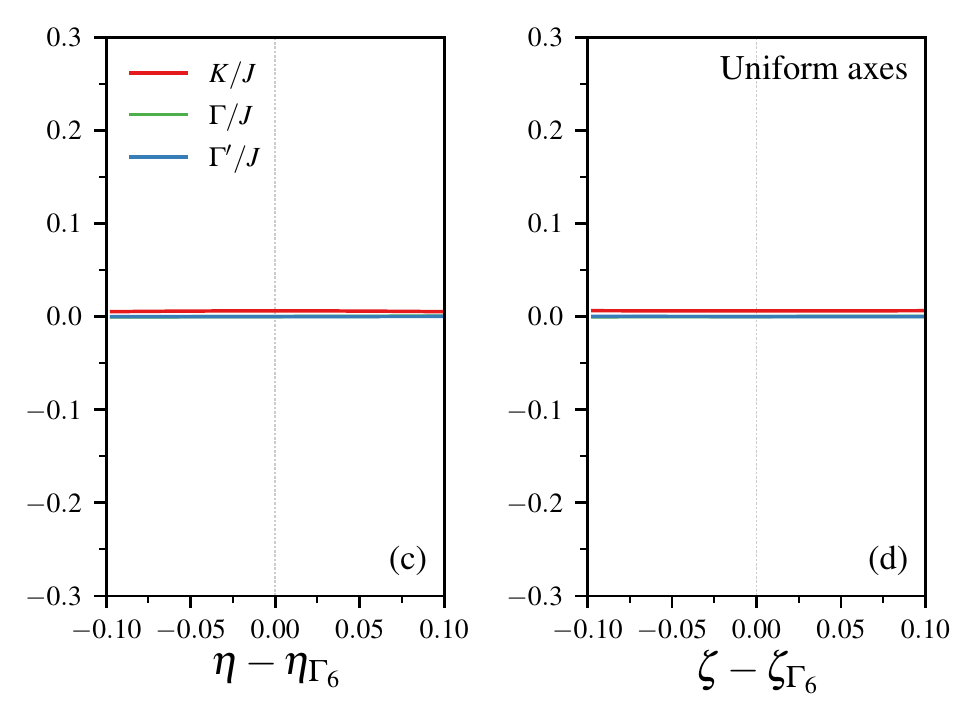}  
  \caption{\label{fig:gamma6-eta-zeta}
    Exchange constants for when moving away from cubic $\Gamma_6$ limit. (a,b) Local axes. Exactly at the $\Gamma_6$ point the \ac{DM} interaction is zero, while away from this point it is the leading sub-dominant exchange. (c,d) Uniform axes. The Heisenberg limit is significantly more robust due to \ac{DM} interaction being forbidden, with the symmetric anisotropic terms developing very weakly upon deviating from the $\Gamma_6$ limit.
  }
\end{figure}

The behavior of the exchanges close to, but away from the cubic limits is important for understanding real materials where trigonal distortions forbid reaching exactly the $\Gamma_6$ or $\Gamma_7$ points. Since the $\Gamma_7$ limit is sensitive to the details of the calculations, and is unlikely to be robust, we only show deviations from the $\Gamma_6$ limit in detail. As shown in Fig.~\ref{fig:gamma6-eta-zeta}, one finds that the most important deviation from the pure Heisenberg antiferromagnet is the \ac{DM} interaction which grows linearly in both $\eta-\eta_{\Gamma_6}$ and $\zeta-\zeta_{\Gamma_6}$. The symmetric anisotropies $K$, $\Gamma$ (and $\Gamma'$ in the uniform case) develop much more slowly. Consequently, the local axes case is qualitatively different than the uniform case where the \ac{DM} interaction is forbidden on symmetry grounds. Note that even in the case of local axes, extremely close to the $\Gamma_6$ limit the symmetric Kitaev interaction is the sub-dominant term, as it does not vanish, as the \ac{DM} interaction does, upon approaching very close to the cubic limits.

\section{General results}
\begin{figure*}[!htpb]
  \centering
  \includegraphics[width=\textwidth]{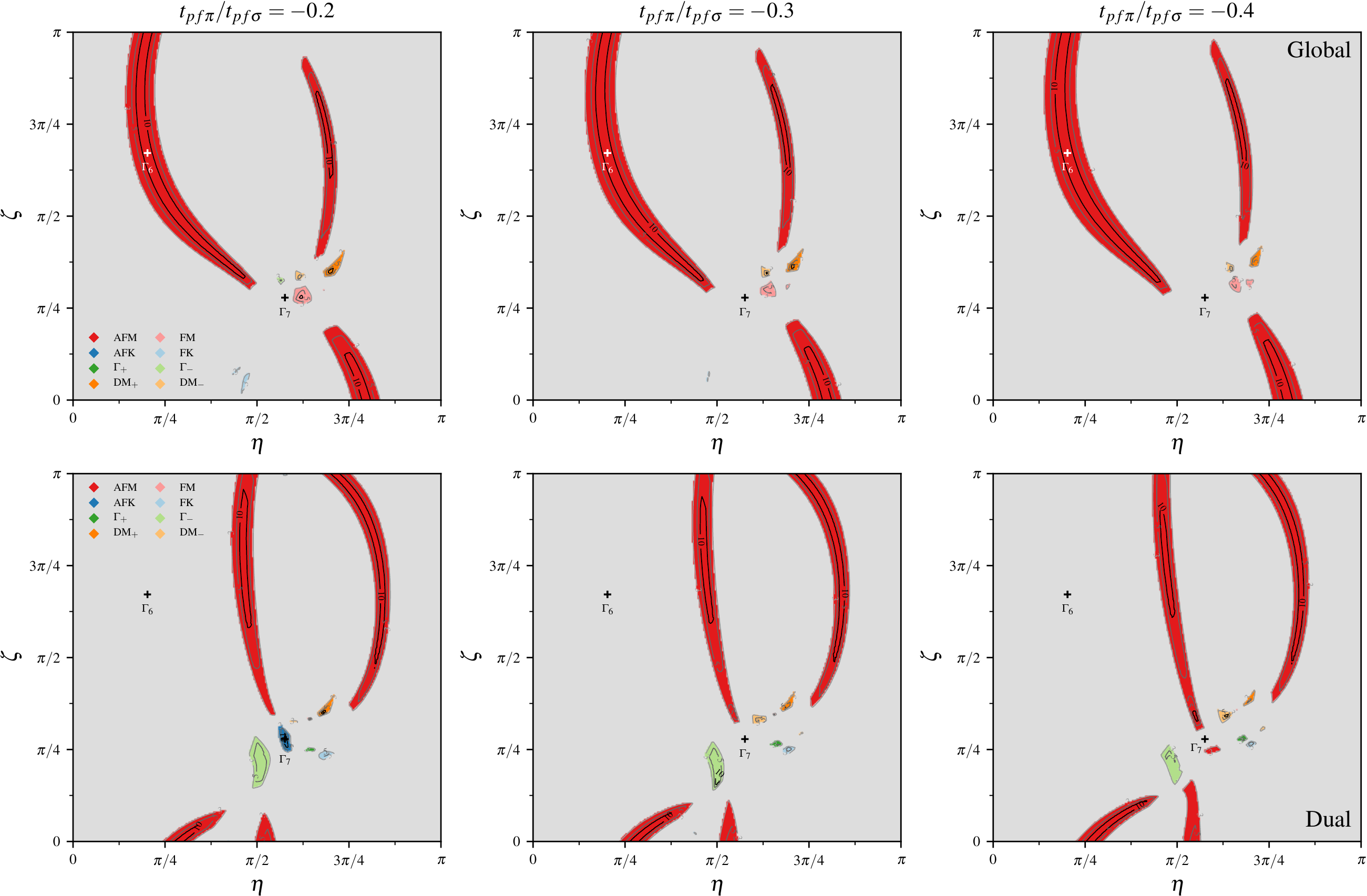}
  \caption{\label{fig:general-non-colinear}
    Exchange regimes for the local axes relevant for the breathing pyrochlore and spinels. If the largest and second largest exchange are less than a factor of three apart in absolute value, we show only gray indicating no clear regime. Otherwise we indicate the dominant exchange ($J$, $K$, $\Gamma$ or $D$) and its sign via a color. Contours of the ratio of the dominant and sub-dominant (second largest) exchange are also shown. Both global and dual representations are shown for three reasonable values of the Slater-Koster ratio $t_{pf\pi}/t_{pf\sigma} = -0.2,-0.3,-0.4$. The legend shows eight possible exchange regimes, depending on which exchange is dominant and its sign: AFM ($J>0$), FM ($J<0$), AFK ($K>0$), FK ($K<0$), $\Gamma_+$ ($\Gamma>0$), $\Gamma_-$ ($\Gamma < 0$), DM$_+$ ($D>0$) and DM$_-$ ($D<0$).
  }
\end{figure*}
\label{sec:general}
We now explore the full parameters space of compositions $(\eta,\zeta)$ for the ground doublet, given that it is unclear how close to the cubic limits (considered in Sec.~\ref{sec:cubic-limit}) the material examples may lie. For simplicity we consider three Slater-Koster ratios: $\rho \equiv t_{pf\pi}/t_{pf\sigma} = -0.2, -0.3, -0.4$ and fix the bond angles to the ideal $90^\circ$. We will show the results in both the equivalent global and dual representations, since the $(J_{zz},J_{\pm},J_{\pm\pm},J_{z\pm})$ representation proves less insightful. For each representation, we rescale the exchanges by the absolute value of the largest exchange, removing any dependence on the overall energy scale set by $t_{pf\sigma}$. The results are presented in Figs.~\ref{fig:general-non-colinear} and \ref{fig:general-colinear}, with the different exchange regimes indicated. We consider the case with local frames (Fig.~\ref{fig:general-non-colinear}) and the case with uniform frames (Fig.~\ref{fig:general-colinear}) in turn.

\subsection{Local frames}
The results for the case of local frames are summarized in Fig.~\ref{fig:general-non-colinear}. Here we see that the nearly perfect Heisenberg antiferromagnetic encountered in the $\Gamma_6$ octahedral limit (see Sec.~\ref{sec:cubic-limit}) extends over a large region in parameter space in both the global and dual parametrizations. As can be clearly seen, these regions are robust and are not strongly affected by variation of the Slater-Koster ratio. While much of the phase space is not in a distinct parameter regime, there are ``islands'' of more pronounced limits close to the $\Gamma_7$ limit. These include near perfect Heisenberg ferromagnets (global) and near perfect antiferromagnetic Kitaev interactions (dual). However, unlike the regions of Heisenberg antiferromagnet, these islands are sensitive to the precise value of the Slater-Koster ratio chosen. For example, the (dual) antiferromagnetic Kitaev point at the $\Gamma_7$ limit is present for $\rho = -0.2$, but is absent for $\rho = -0.3$ and $\rho = -0.4$. Similar appearance and disappearance of these islands as a function of $\rho$ can also be seen for the (dual) islands of dominant \ac{DM} interactions and dominant $\Gamma$ interaction. Due to this dependence on the detailed parameter choices made, which are heretofore unknown, we can see that our predictions for the exchanges near the $\Gamma_7$ point are likely to be significantly less reliable than those near the more robust $\Gamma_6$ region (as expected from the considerations of Sec.~\ref{sec:cubic-limit}).

\subsection{Uniform frames}
\begin{figure*}[!htpb]
  \centering
  \includegraphics[width=\textwidth]{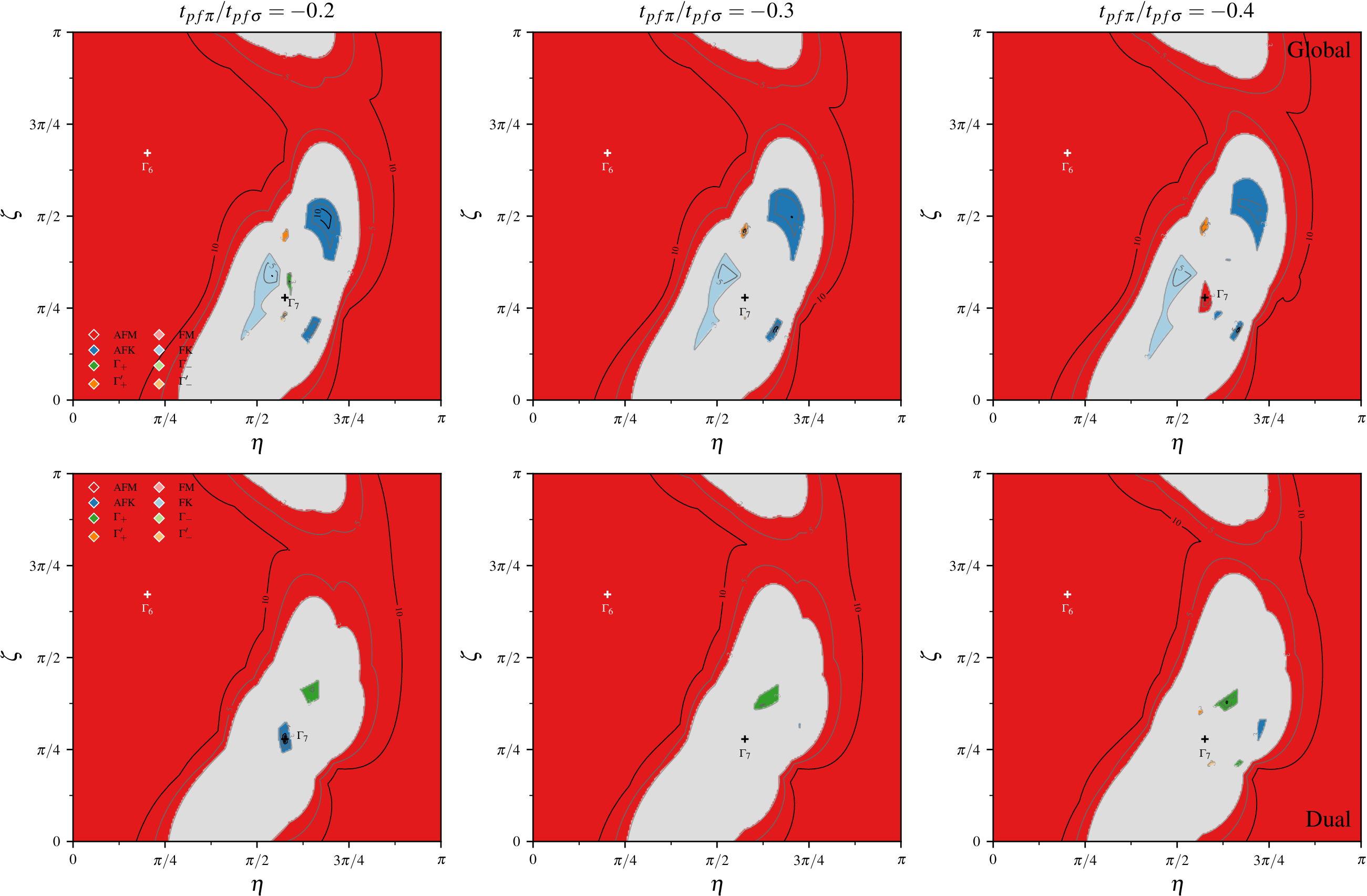}
  \caption{\label{fig:general-colinear}
    Exchange regimes for the uniform axes relevant for triangular or honeycomb compounds. If the largest and second largest exchange are less than a factor of three apart in absolute value, we show only gray indicating no clear regime. Otherwise we indicate the dominant exchange ($J$, $K$, $\Gamma$ or $\Gamma'$) and its sign via a color. Contours of the ratio of the dominant and sub-dominant (second largest) exchange are also shown. Both global and dual representations are shown for three reasonable values of the Slater-Koster ratio $t_{pf\pi}/t_{pf\sigma} = -0.2,-0.3,-0.4$.The legend shows eight possible exchange regimes, depending on which exchange is dominant and its sign: AFM ($J>0$), FM ($J<0$), AFK ($K>0$), FK ($K<0$), $\Gamma_+$ ($\Gamma>0$), $\Gamma_-$ ($\Gamma < 0$), $\Gamma'_+$ ($\Gamma'>0$) and $\Gamma'_-$ ($\Gamma'<0$).
  }
\end{figure*}
The result for the uniform case are summarized in Fig.~\ref{fig:general-non-colinear}. As in the non-uniform case, the $\Gamma_6$ limit is embedded in a robust region of antiferromagnetic  Heisenberg interactions. This occupies a significantly larger region of parameter space here than in the case with local axes. This arises since the antisymmetric \ac{DM} interaction is now forbidden and, like in the case with local axes, the symmetric anisotropies only develop weakly as one moves away from the $\Gamma_6$ point. Similar to the case with local axes, the region around the $\Gamma_7$ limit also hosts other anisotropic regimes, in this case both ferro- and antiferromagnetic Kitaev limits in the global basis and an antiferromagnetic Kitaev limit in the dual basis. Both of these lie somewhat off of the pure $\Gamma_7$ limit and are somewhat sensitive to changes in the Slater-Koster ratio, though less so than in the case with local axes. Also present for some Slater-Koster ratios are (global) ferromagnets and dominant (dual) $\Gamma > 0$ interactions.

\section{Applications to materials}
\label{sec:applications}
We now apply the general results of Sec.~\ref{sec:general} to some specific ytterbium based rare-earth magnets.
\subsection{Spinels}
\label{sec:spinels}
The rare-earth chalcogenide spinels, \spinel{A}{R}{X} share many structural features with the breathing pyrochlore \byzo{}. Thus, given the success of these calculations in reproducing those exchanges (see Sec.~\ref{sec:validation}), we expect the methods of Sec.~\ref{sec:super-exchange} to work reasonably well for the spinels. These compounds have space group $Fd\cb{3}m$ (no. 227) with the rare-earth R (Wyckoff site $16d$) forming a pyrochlore lattice and the A ion (Wyckoff site $8a$) being non-magnetic. The ligand, denoted as X (Wyckoff site $32e$), forms distorted octahedra around the rare-earths. The ligand position, which we denote as $x$, varies from material to material, but does not stray too far~\cite{lau2005spinel} from $x = 1/4$ which yields ideal X octahedra. These RX\tsub{6} octahedra are joined together in an edge-sharing network as illustrated in Fig.~\ref{fig:structures}.

Unfortunately, there does not appear to have been direct measurements of the crystal field spectrum, and thus the composition of the ground doublet, for any of the ytterbium spinels. While several estimates~\cite{spinel-cef-1,spinel-cef-2,pawlak1988magnetic,higo-2016-frustrated} exist in the literature, they are based on fitting of the magnetic susceptibility and use restricted, mostly cubic, ansatzes for the crystal field interaction parameters. These thus produce nearly perfect $\Gamma_6$ ground state doublets and thus essentially reproduce the results of Sec.~\ref{sec:cubic-limit}.  Given the ideal $\Gamma_6$ limit hosts a near perfect Heisenberg antiferromagnet, the physics is highly sensitive to any sub-dominant perturbations.

\begin{figure}[tp]
  \centering
  \includegraphics[width=0.9\columnwidth]{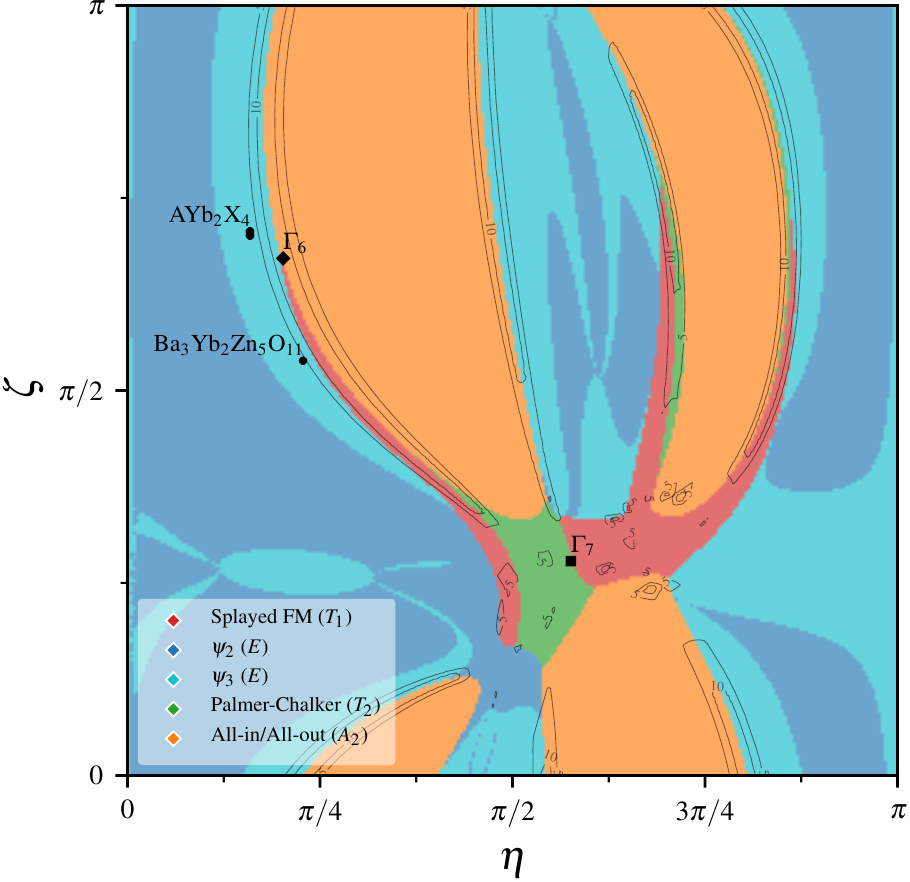}\hspace{0.5cm}
  \caption{\label{fig:spinels-phases}
Semi-classical phase diagram for exchange constants computed for the spinel structure with a bond angle of $93.6^{\circ}$ and Slater-Koster ratio of $t_{pf\pi}/t_{pf\sigma} = -0.3$.  The classical energy is minimized over possible $\vec{k}=0$ ground states~\cite{yan2017anisotropic}, with the accidental degeneracy in the $\Gamma_5$ ($E$) phase resolved by $1/S$ corrections of linear spin-wave theory~\cite{zhitomirsky2012obd,savary2012obd}. Contours show the combined relative ratios of the sub-dominant to dominant exchange (for global and dual parametrizations) as discussed in Sec.~\ref{sec:super-exchange} and as used in Figs.~\ref{fig:general-non-colinear} and \ref{fig:general-colinear}.
  }
\end{figure}

From the considerations of Sec.~\ref{sec:cubic-limit}, we expect a subdominant \ac{DM} interaction to exist, with the qualitative features depending on its sign. In the full phase diagram of the anisotropic exchange model of Eq.~(\ref{eq:model}) one generically expects four magnetically ordered phases with zero wave-vector: an \ac{AIAO} state, a \ac{PC} state, a \ac{SFM} state or a $\Gamma_5$ state~\cite{yan2017anisotropic}. To illustrate this, we have computed the classical ground state for the exchanges predicted for each crystal field composition, as shown in Fig.~\ref{fig:spinels-phases}. For a Heisenberg antiferromagnet with small direct \ac{DM} interactions ($D>0$) one expects an \ac{AIAO} state~\cite{chern2010pyrochlore}. For small indirect \ac{DM} interaction ($D<0$) the situation is more complex as this is phase boundary between the \ac{SFM} and $\Gamma_5$ states when the symmetric anisotropies are included~\cite{chern2010pyrochlore,elhajal2005,canals2008,yan2017anisotropic,wong2013xy}. These select the \ac{SFM} state when $K + \Gamma >0$ and the $\Gamma_5$ states when $K +\Gamma <0$.  The physics along the boundary with $K=\Gamma=0$ is more involved and includes an additional one-dimensional degenerate manifold of states along with the $\Gamma_5$ states~\cite{chern2010pyrochlore,elhajal2005,canals2008}. In Fig.~\ref{fig:spinels-phases}, we see that, for the indirect case, mostly the $\Gamma_5$ state is selected by the subdominant $K$ and $\Gamma$ exchanges, save for a small window of \ac{SFM} emerging from the pure octahedral limit. We thus see that it is natural for the spinels to have either a $\Gamma_5$ or \ac{AIAO} ground state, given some unknown deviations from the ideal $\Gamma_6$ limit. Within the $\Gamma_5$ manifold, the ground state has an accidental continuous degeneracy~\cite{savary2012obd,zhitomirsky2012obd} which will be lifted through a number of competing order-by-disorder mechanisms~\cite{zhitomirsky2012obd,savary2012obd,oitmaa2013obd,rau-2016-order}. For the pure nearest-neighbor model of Eq.~(\ref{eq:model}) at zero temperature, the leading effect is quantum order-by-disorder~\cite{savary2012obd,zhitomirsky2012obd} which selects either the non-colinear $\psi_2$ or colinear $\psi_3$ state.  Within the $\Gamma_5$ regions, we have computed the zero-point energies and show the state selected by $1/S$ corrections in the usual linear spin-wave theory~\cite{savary2012obd}. We note that \emph{both} $\psi_2$ and $\psi_3$ states are found relatively close to the cubic $\Gamma_6$ limit, with $\psi_3$ appearing immediately adjacent.

To make a more detailed prediction, we need some estimate of the spectral composition of the crystal field ground doublet. To estimate the crystal fields in the ytterbium spinels, we leverage the detailed analysis carried for the crystal field of the related \spinel{Mg}{Er}{Se} spinel.~\cite{reig-2017-spin}.  Through fitting to the results of inelastic neutron scattering, it is found that the crystal field parameters are somewhat different than expected for an approximately cubic crystal field. In addition, the trigonal contributions are found not to be well described by a point-charge model including only the nearest-neighbor ligands~\footnote{ This failure of the point charge model could be due to covalency effects due to the larger $4p$ orbitals of the Se ligands or a number of non-electrostatic effects~\cite{newman2007crystal}.}.  Similar results~\cite{lau2005spinel} have been found for \spinel{Cd}{Er}{S} though, as in the ytterbium spinels, only bulk probes (in this case the magnetization) were analyzed.  Given these parameters, we estimate the crystal field for the ytterbium spinels by rescaling the fitted parameters found for \spinel{Mg}{Er}{Se}~\cite{reig-2017-spin}.  Such a rescaling procedure was used in Ref.~[\onlinecite{bertin2012crystal}] and was been found to be relatively accurate across the full series of rare-earth pyrochlore titanates. In this procedure, the crystal field parameters, denoted as $B_{kq}$, for the \spinel{A}{Yb}{X} are determined from the $B^0_{kq}$ relevant for \spinel{Mg}{Er}{Se} via
\begin{equation}
  \label{eq:rescaling}
  B_{kq} \equiv
  \frac{\theta^{(k)}_{} \avg{r^k}_{}}{\theta^{(k)}_{0} \avg{r^k}^{}_{0}}
  \left(\frac{a}{a_0}\right)^{-(k+1)} B^{0}_{kq},
\end{equation}
where the $\theta^{(k)}$, $\avg{r^k}$ and $\theta^{(k)}_0$, $\avg{r^k}^{}_0$ are the Stevens' parameters~\cite{stevens1952matrix} and radial integrals~\cite{freeman1979dirac} for Yb and Er respectively, and $a$ and $a_0$ are the lattice constants of the target \spinel{A}{Yb}{X} spinel~\cite{higo-2016-frustrated} and \spinel{Mg}{Er}{Se}~\cite{reig-2017-spin}. We are able to reproduce the results of Ref.~[\onlinecite{reig-2017-spin}] using the parameters~\footnote{ Note that the sign of the $B_{43}$ and $B_{63}$ cannot be determined by an examination of powder inelastic neutron scattering data. For most probes of interest this is immaterial, but it is relevant for our super-exchange calculations (see Sec.~\ref{sec:single-ion} for a discussion of this distinction). We have picked the sign such that the ground doublet is as close as possible to the octahedral $\Gamma_6$ limit.}, translated to the conventional notation
\begin{align}
  B^{0}_{20} &= -4.227\cdot 10^{-2} \meV, &  B^{0}_{40} &= -6.116 \cdot 10^{-4} \meV, \nonumber \\
  B^{0}_{43} &= -1.338\cdot 10^{-2}\meV, &  B^{0}_{60} &= +3.315 \cdot 10^{-6} \meV, \nonumber \\
  B^{0}_{63} &= -3.840\cdot 10^{-5}\meV, &  B^{0}_{66} &= +2.266 \cdot 10^{-5} \meV. \nonumber   
\end{align}
While this procedure is likely to be most accurate for the selenides \spinel{Cd}{Yb}{Se} and \spinel{Mg}{Yb}{Se} given the common ligand, we will apply to the sulphides as well. Note that this procedure ignores variations in the ligand structural parameter $x$.  While this parameter is only reported in the literature for \spinel{Cd}{Yb}{S} (with $x \sim 0.2579$~\cite{lau2005spinel} and $x\sim 0.2594$~\cite{long-range-2017}) and for \spinel{Cd}{Yb}{Se} (with $x \sim 0.2575$~\cite{long-range-2017}), we note that this parameter does not appear to vary strongly upon variation of of A = Mg, Cd or the ligand X = S, Se ~\cite{lau2005spinel,reig-2017-spin,long-range-2017}. With these rescaled parameters, one finds energy levels that are broadly consistent with that found from fitting the magnetic susceptibility~\cite{higo-2016-frustrated} for \spinel{Cd}{Yb}{S} and \spinel{Mg}{Yb}{S}, as presented in Table~\ref{tab:spinel-data}.
\begin{table*}[tp]
  \begin{ruledtabular}
    \begin{tabular}{ccccccccccc}
      Spinel & $a$ [\AA] & $\theta_{\rm CW}$ [K] &  $T_N$ [K]
      & $E_1$ [meV] & $E_2$ [meV] & $E_3$ [meV]
      & $g_z$ & $g_{\pm}$ 
      & $\eta$ & $\zeta$                                                           
      \\
  \hline
      \spinel{Cd}{Yb}{S}~\cite{higo-2016-frustrated} & $11.075$& $-10.0$  & $1.8$
      & $23.47$ & $33.46$ & $63.69$
      & $-3.587$ & $-2.188$
      & $0.4995$ & $2.206$
      \\
      \spinel{Mg}{Yb}{S}~\cite{higo-2016-frustrated}& $10.972$  & $-10.4$  & $1.4$
      & $24.83$ & $35.06$ & $66.67$
      & $-3.591$ & $-2.185$
      & $0.4996$ & $2.203$
      \\
      \spinel{Cd}{Yb}{Se}~\cite{higo-2016-frustrated}& $11.539$ & $-9.3$  & $1.7$
      & $18.32$ & $27.28$ & $52.14$
      & $-3.560$ & $-2.206$                            
      & $0.4999$ & $2.222$
      \\      
      \spinel{Mg}{Yb}{Se}~\cite{higo-2016-frustrated}& $11.464$  & $-9.2$  & $1.4$
      & $19.06$ & $28.17$ & $53.82$
      & $-3.564$ & $-2.204$
      & $0.4999$ & $2.219$ \\ \hline
      \spinel{Cd}{Yb}{S}~\cite{long-range-2017} & $11.003$& $-13.0$  & $1.92$
&&& &&
      \\
      \spinel{Cd}{Yb}{Se}~\cite{long-range-2017} & $11.455$& $-11.0$  & $1.75$
&&& &&                                                                             
    \end{tabular}
    \caption{\label{tab:spinel-data}
      Survey of some experimental data on ytterbium spinels, including lattice constant~\cite{higo-2016-frustrated,long-range-2017}, Curie-Weiss temperature~\cite{higo-2016-frustrated,long-range-2017} and N\'eel temperature~\cite{higo-2016-frustrated,long-range-2017}. We show the excited crystal field energy levels, $g$-factors and ground doublet composition computed for the \spinel{A}{Yb}{X} spinels using the crystal
      structures from Ref.~[\onlinecite{higo-2016-frustrated}]. The crystal field parameters were obtained by rescaling from the fitted parameters [Eq.~(\ref{eq:rescaling})] for \spinel{Mg}{Er}{Se} found in Ref.~[\onlinecite{reig-2017-spin}]. For all spinels we assume a Yb-X-Yb bond angle of $\sim 93.6^\circ$ as found in \spinel{Cd}{Yb}{S}~\cite{lau2005spinel}.
    }    
  \end{ruledtabular}
\end{table*}
The resulting ground doublet for these crystal field is relatively close to a $\Gamma_6$ doublet, with $(\eta,\zeta) \sim (0.5,2.2)$ for all compounds (see Table~\ref{tab:spinel-data}). These parameters yield weakly Ising-like $g$-factors with $(g_z,g_{\pm}) \sim (-3.6,-2.2)$. While still quite close to the $\Gamma_6$ limit, this composition is significantly further away than what is obtained from the crystal field parameters reported in Ref.~[\onlinecite{higo-2016-frustrated}].  The predicted exchange constants lie in the regime of strong antiferromagnetic Heisenberg coupling with sub-dominant \ac{DM} interactions near the $\Gamma_6$ point (see Sec.~\ref{sec:cubic-limit}). In the global basis one finds
\begin{align}
  \label{eq:spinel-exchange}
  K/J &\sim -0.03,&   \Gamma/J &\sim -0.02 ,&   D/J &\sim -0.3,
\end{align}
similar to what is found in \byzo{} (see Sec.~\ref{sec:validation}). These exchanges place the spinels into a region with a classical $\Gamma_5$ ground state, with semi-classical $1/S$ corrections selecting $\psi_3$ ordering. The presence of a large, positive $J$ and subdominant, indirect \ac{DM} interaction are not sensitive to small changes in the bond angle or Slater-Koster ratio. This is also true of the selection of the $\Gamma_5$ states over the \ac{SFM} state, in spite of the much smaller scale of these terms. We do note, however, that if we assume the overall energy scale is of order the Curie-Weiss temperature, then the small symmetric exchanges are of the same order as those expected from magnetic dipole interactions, as discussed in App.~\ref{app:dipolar:spinels}. Since we are close to a phase boundary controlled by the sum, $K+\Gamma$, of these subdominant terms, we must consider these contributions carefully.  However, one has that the contributions to the sum $K+\Gamma$ from the nearest-neighbor part of the dipolar interaction approximately cancel and thus does not affect the selection of the ground state.  We note that the state selected via order-by-thermal-disorder near the ordering temperature can be in principle different that that selected by quantum or thermal fluctuations near $T=0$~\cite{mcclarty2014,jav2015}. Indeed, if we consider $K=\Gamma=0$, thermal fluctuations select the $\psi_2$ state near $T_N$, but $\psi_3$ near $T=0$ (classically)~\cite{chern2010pyrochlore,elhajal2005,canals2008}. Whether such an intermediate $\psi_2$ phase would be present for the parameter regime of interest would likely depend on the precise values of $K$, $\Gamma$ and how strongly the quantum selection competes with thermal selection.

Experimentally, one finds that each of the spinel compounds has an antiferromagnetic Curie-Weiss constant of $\sim 9-10 \K$~\cite{higo-2016-frustrated}, roughly consistent with the exchange scale found in \byzo{}~\cite{rau-2016-byzo}. At low temperatures, each of the four compounds orders antiferromagnetically, with N\'eel temperatures in the range $T_{\rm N} \sim 1.4-1.8\K$~\cite{higo-2016-frustrated}, somewhat strongly reduced from the na\"ive energy scale of $10\K$~\cite{higo-2016-frustrated}.  Below $T_N$, the specific heat was found to decrease roughly as $\sim T^3$ as $T \rightarrow 0$, suggesting the presence of linearly dispersing gapless modes~\cite{higo-2016-frustrated}. Evidence for such gapless excitations in \spinel{Cd}{Yb}{S} has also been suggested from electron spin-resonance (ESR) measurements~\cite{yoshizawa2015high}. Of the possible ordered states expected for the nearest-neighbor anisotropic exchange model, this is \emph{only} consistent with the (nearly) gapless spectrum expected in a $\Gamma_5$ state~\footnote{While generically the other states (i.e. the \ac{SFM}, \ac{AIAO} or \ac{PC} states) would have a large gap (and thus would be inconsistent with the specific heat data), this gap could be rather small if one is close to a phase boundary. It thus could be difficult to distinguish the two behaviors in practice (a similar situation has been proposed for \abo{Gd}{Sn}~\cite{quilliam2007gso}).}, as found in \eto{}~\cite{champion2003obd,savary2012obd,ross2014obd,petit2014}. Indeed the
presence of $\Gamma_5$ order was recently directly observed in the \spinel{Cd}{Yb}{S} and \spinel{Cd}{Yb}{Se} spinels by neutron
diffraction~\cite{long-range-2017}. Given these considerations, the \spinel{A}{Yb}{X} may represent a particular clean example of quantum order-by-disorder, free from some of the potential complications of the \abo{Er}{M} family~\cite{rau-2016-order}, similar to what has been proposed for \abo{Yb}{Ge}~\cite{petit2014,jaubert2015multiphase,petit2015dynamics}.

However, there are some key differences between these compounds, in particular in the muon spin resonance ($\mu$SR) signals for the \spinel{Cd}{Yb}{S} and \spinel{Mg}{Yb}{S} compounds~\cite{higo-2016-frustrated}. One finds that the Cd compounds show well-defined~\cite{higo-2016-frustrated,long-range-2017} oscillations in the mean asymmetry below $T_{\rm N}$, as expected for an ordered state, while one of the Mg compounds shows no such oscillations~\cite{higo-2016-frustrated}. This  absence of static internal field at the muon site in \spinel{Mg}{Yb}{S} was interpreted  in Ref.~[\onlinecite{higo-2016-frustrated}] as a possible signature of incommensurate order. However, given that the muon implantation site is unknown and, further, the muon can distort the crystal environment, it is unclear how directly one can interpret these results. Further, a lack of pronounced oscillations has also been seen in \abo{Er}{Ti}~\cite{lago2005magnetic} which has an uncontroversial $\psi_2$ ground state~\cite{poole2007magnetic} or in \abo{Yb}{Ge}~\cite{hallas2016musr}, where a $\psi_3$ ground state may be expected~\cite{jaubert2015multiphase}. In addition to the difference in the $\mu$SR signals, there is also a difference in the field-cooled vs. zero field-cooled magnetic susceptibility, with \spinel{Cd}{Yb}{S} showing a bifurcation at $T_{\rm N}$ while \spinel{Mg}{Yb}{S} does not~\cite{higo-2016-frustrated}. These experiments are broadly consistent with the scenario outlined where a $\Gamma_5$ state is the ground state. The simplest explanation may be that the ground state is $\Gamma_5$, with perhaps a differences in proximity to the phase boundary with the \ac{SFM} state~\cite{jaubert2015multiphase,yan2017anisotropic} accounting for the differing $\mu$SR and susceptibility measurements between \spinel{Cd}{Yb}{S} and \spinel{Mg}{Yb}{S}

On a more phenomenological level, one can look at the trends found in the \spinel{A}{Yb}{X} series as one varies the ligand, X, and the A ion (see Table~\ref{tab:spinel-data}). One finds that while the Curie-Weiss ($\theta_{\rm CW}$) temperature varies mostly with the choice of ligand~\cite{higo-2016-frustrated}, both $T_N$ and the spin-wave velocity extracted from the specific heat follow the choice of A = Cd or Mg~\cite{higo-2016-frustrated}. One possibility is that the lattice constant, which follows mostly with the choice of ligand, determines the overall exchange scale (and thus $\theta_{\rm CW}$), while $T_N$ is determined by subdominant exchanges and thus details of the ground doublet composition. Since we obtained our ground doublet composition from a rescaling of the parameters found for the Mg based spinel, \spinel{Mg}{Er}{Se}, the crystal field could be somewhat different for a Cd based spinel, changing some of the details of the exchanges. We find that if one rescales the fitted crystal field parameters from Ref.~[\onlinecite{gao2017dipolar}] for the \spinel{Cd}{Er}{S} or \spinel{Cd}{Er}{Se} spinels, the results are inconclusive; there are small changes in the exchanges, but they do not follow the trends described above. Whether this is a failure of our super-exchange calculation to capture these fine details, or whether this is due to uncertainties in the rescaling procedure used obtain the crystal parameters remains to be seen.~\footnote{We note that there is some ambiguity in fitting the crystal field levels of a \rth{Yb} ion from only powder inelastic neutron data. Since there are four doublets, one has only five pieces of information: three energy differences and two relative intensities (assuming the overall intensity is not measured on an absolute scale). Since the crystal field Hamiltonian has six real parameters when one considers trigonal symmetry, the problem is somewhat ill-posed. To resolve this one must include additional data (e.g. susceptibility or the behavior of the levels in a magnetic field) or measure the overall intensity scale of the neutron scattering experiment.
}

Finally, it is interesting to note that these exchanges are not too far from those found in a recent study~\cite{thompson2017quasiparticle} of \yto{}. In the local basis, the exchanges of Eq.~(\ref{eq:spinel-exchange}) have dominant $J_{z\pm} > 0$ with
\begin{align*}
  J_{zz}/J_{z\pm} &\sim -0.13, &   J_{\pm}/J_{z\pm} &\sim +0.48, &   J_{\pm\pm}/J_{z\pm} &\sim +0.53,  
\end{align*}
quite similar, keeping in mind that the sign of $J_{z\pm}$ can be changed by a local spin rotation. Indeed, if one recasts the exchanges of Ref.~[\onlinecite{thompson2017quasiparticle}] for \yto{} to the dual language of Eq.~(\ref{eq:pyro:dual}), then one has $\tilde{J}>0$ with
\begin{align}
  \tilde{K}/\tilde{J} &= 0.01 , & \tilde{\Gamma}/\tilde{J} &= +0.1, &   \tilde{D}/\tilde{J} &= -0.5,
\end{align}
that is a large indirect \ac{DM} interaction with (relatively) small symmetric anisotropies.  One can also note here that $\tilde{K} + \tilde{\Gamma} > 0$, selecting the \ac{SFM} state classically, as (currently)
expected for \yto{}~\cite{thompson2017quasiparticle}. While the physics of \yto{} has been proposed to be related to its proximity to the $\Gamma_5$-\ac{SFM} boundary~\cite{jaubert2015multiphase,petit2015dynamics}, it becomes particularly clear in this dual language where it connects smoothly to the work of Refs.~[\onlinecite{chern2010pyrochlore,elhajal2005,canals2008}] where the $\Gamma_5$-SFM boundary is obtained by fixing $K=\Gamma=0$ and varying $D/J < 0$.

We thus conclude that the physics of the \spinel{A}{Yb}{X} spinels may be closely tied to that of the \abo{Yb}{M} pyrochlores.  These parallels are also manifest experimentally; for example, a similar double-peak structure in the specific heat, as observed in the \abo{Yb}{M} (M = Ti, Ge, Sn) compounds~\cite{hallas2016universal}, is also present in the \spinel{A}{Yb}{X} spinels~\cite{higo-2016-frustrated,long-range-2017}. Given that the strong competition~\cite{wong2013xy,jaubert2015multiphase,yan2017anisotropic} between between nearby $\Gamma_5$ (found in \abo{Yb}{Ge}~\cite{dun2015ybgeo,hallas2016xy}) and \ac{SFM} phases (found in \abo{Yb}{Ti} and \abo{Yb}{Sn}~\cite{dun2014chemical}) may be responsible for some of the physics of the \abo{Yb}{M} family, it may be worthwhile to explore whether some of the exotic dynamical properties~\cite{petit2015dynamics,hallas2016universal,thompson2017quasiparticle} found in (for example in \abo{Yb}{Ge}) may carry over to the \spinel{A}{Yb}{X} spinels. We see this as a highly promising avenue for future experimental investigations.

\subsection{Triangular}
\label{sec:triangular}
We now consider the triangular compound \ybmggao{} which has recently attracted attention as a potential quantum spin liquid candidate~\cite{li-2015-rare-earth}. Here, the \rth{Yb} ions form a triangular lattice, supported by a network of edge-sharing oxygen octahedra (see the idealized form in Fig.~\ref{fig:structures}). The bond angle in this compound is the furthest from ideal we consider, being close to $99^\circ$~\cite{li-2015-rare-earth}.  Additionally, there is the complication of chemical disorder, with the Mg$^{2+}$ and Ga$^{3+}$ ions not forming a periodic structure~\cite{li-2015-rare-earth}. Experimentally, one finds no magnetic ordering down to $\sim 60 \mK$~\cite{li-2015-rare-earth}, well below the expected magnetic energy scale of $\sim 1\K$~\cite{li-2015-rare-earth}. At the lowest temperatures, the specific heat follows a power law $\sim T^{0.7}$ suggesting gapless excitations~\cite{li_gapless_2015}.  The excitation spectrum, as probed by inelastic neutron scattering, is consistent with this, showing a broad, gapless continuum with few distinct features as a function of energy or momentum~\cite{shen_evidence_2016,paddison_continuous_2016}. These characteristics have been interpreted as evidence for a gapless quantum spin liquid ground state in this compound~\cite{shen_evidence_2016}.

While promising, there have been several challenges to the interpretation as a quantum spin liquid.  First is the absence of magnetic thermal conductivity at low temperatures~\cite{xu_absence_2016}.  This is at odds with the expectation that in a quantum spin liquid with gapless excitations, as seen in the specific heat and in neutron scattering, should transport heat~\cite{shen_evidence_2016,paddison_continuous_2016}. Second, and perhaps more importantly, is the role the disorder on the Mg and Ga sites affects the magnetic physics~\cite{li-2015-rare-earth}. It has been suggested that the experimental data shows evidence for a distribution of $g$-factors for the Yb spins due to the Mg/Ga mixing~\cite{li_crystalline_2017}. Further, it has been argued that certain kinds of exchange disorder could mimic some of the features that have been interpreted as evidence for a quantum spin liquid ground state~\cite{zhu_disorder-induced_2017}.

We first address what is known of the (suitably disordered averaged) exchange constants. The work of Ref.~[\onlinecite{li-2015-rare-earth}] first addressed this question, obtaining values for all four exchanges; in our notation these parameters read
\begin{align*}
  \label{eq:ybmggao-ex}
  J_{zz} &= +0.98 \K, &   J_{\pm} &= -0.90 \K, \\   J_{\pm\pm} &= \pm 0.15 \K, &   J_{z\pm} &= \pm 0.04 \K,
\end{align*}
where the signs of $J_{\pm\pm}$ and $J_{z\pm}$ were undetermined~\cite{li-2015-rare-earth}. These exchanges were obtained by a sequence of experimental fits: first the $g$-factors were extracted from high-field magnetization, then $J_{zz}$ and $J_{\pm}$ from the Curie-Weiss constants and finally $J_{\pm\pm}$ and $J_{z\pm}$ from ESR linewidths~\cite{li-2015-rare-earth}. However, the first two steps in this process, upon which the last relies, vary somewhat in the literature. For example, the Curie-Weiss constants found in Ref.~[\onlinecite{shen_evidence_2016}] imply exchanges of $J_{zz} = 2.13\K$ and $J_{\pm} = -1.59\K$. The estimates of $J_{zz}/J_{\pm}$ obtained in Ref.~[\onlinecite{paddison_continuous_2016}] through fitting to inelastic neutron scattering in field and in Ref.~[\onlinecite{toth_strong_2017}] through fitting to elastic diffuse neutron scattering also differ somewhat from what was found in Ref.~[\onlinecite{li-2015-rare-earth}]. Given these results, it seems unclear what exchanges are reasonable for \ybmggao{}, with the only common thread being that $J_{zz} \sim O(1\K)$ and $J_{\pm} < 0$ with $J_{zz} \gtrsim |J_{\pm}|$.

Given this uncertainty, in this section we will apply the exchange calculations developed in this work to \ybmggao{}.  Structurally, this compound has the same pattern of (approximately) edge-sharing YbX\tsub{6} octahedra found in the breathing pyrochlore and the spinels. The results of Sec.~\ref{sec:general} for the case with uniform axes can thus be applied, once the larger bond angle of~\cite{li-2015-rare-earth} (approximately) $\sim 99^{\circ}$ is taken into account in the Slater-Koster overlaps of Eq.~(\ref{eq:hyb}). As in the case of spinels, determining the exchange regime relevant for \ybmggao{} rests on an accurate determination of the composition of the crystal field ground doublet.

Estimates ignoring the Mg/Ga disorder (based on the saturation of the magnetization in high-field) give a weak Ising anisotropy, with $(g_z,g_{\pm}) \sim (+3.721, -3.060)$, with only a single combination of signs realizable within a strict $\Gamma_4$ doublet.  Including the Ga/Mg site disorder that locally modifies the crystal field is a complex problem~\cite{li_crystalline_2017}. We follow Ref.~[\onlinecite{li_crystalline_2017}] and consider an ensemble of possible crystal field environments for the Yb ion, restricting to seven distinct Mg/Ga configurations. This set of crystal field compositions and their associated $g$-factors are listed in Table~\ref{tab:triangular-cef}. We see that, for each of these crystal field parameters, one finds that $g_z > 0$ and $g_{\pm} < 0$, as found for parameter set which assumed no significant disorder~\cite{li-2015-rare-earth}. It is suggested in Ref.~[\onlinecite{li_crystalline_2017}] that this modification of the crystal field due to the Mg/Ga disorder is not primarily due to the direct effects of the charge disorder, but due to its distortion of the oxygen cage and off-centering of the Yb ion. This distortion modifies both the distances and angles of the oxygen ligands relative to the crystal axes~\cite{li_crystalline_2017}, and thus affects the exchanges through both the ground doublet composition and through changes in the ligand bond angles.
\begin{table}[tp]
  \begin{ruledtabular}
    \begin{tabular}{rcccc}
      Env. & $\eta$ & $\zeta$ & $g_z$ & $g_{\pm}$ \\ \hline
      $1$ & $1.197$ & $2.440$ &  $+3.697$ & $-3.221$ \\
      $2$ & $1.195$ & $2.439$ &  $+3.670$ & $-3.229$ \\
      $3$ & $1.218$ & $2.444$ &  $+3.873$ & $-3.163$ \\
      $4$ & $1.147$ & $2.423$ &  $+3.213$ & $-3.352$ \\
      $5$ & $1.242$ & $2.449$ &  $+4.066$ & $-3.092$ \\
      $6$ & $1.144$ & $2.422$ &  $+3.182$ & $-3.359$ \\
      $7$ & $1.179$ & $2.434$ &  $+3.530$ & $-3.270$ \\
    \end{tabular}
    \caption{\label{tab:triangular-cef}
      Computed crystal field ground doublets and $g$-factors for the
seven Mg/Ga local environments and crystal field parameters proposed
in Ref.~[\onlinecite{li_crystalline_2017}].
    }
  \end{ruledtabular}
\end{table}

\begin{table}
  \begin{ruledtabular}
    \begin{tabular}{cccccc}     
      $\theta$ & Env. & $J_{\pm}/J_{zz}$ & $J_{\pm\pm}/J_{zz}$ & $J_{z\pm}/J_{zz}$ & $J_{zz}/J_{zz}^0$\\
      \hline
      $97^\circ$
& 1 & $-0.44$ & $-0.06$ & $0.03$ & $1.63$ \\
& 2 & $-0.44$ & $-0.06$ & $0.03$ & $1.65$ \\
& 3 & $-0.44$ & $-0.07$ & $0.04$ & $1.49$ \\
& 4 & $-0.44$ & $-0.06$ & $0.02$ & $2.0$ \\
& 5 & $-0.44$ & $-0.07$ & $0.04$ & $1.34$ \\
& 6 & $-0.44$ & $-0.06$ & $0.02$ & $2.02$ \\
& 7 & $-0.44$ & $-0.06$ & $0.03$ & $1.76$ \\
      \hline
      $99^\circ$
& 1 & $-0.36$ & $-0.14$ & $0.04$ & $1.0$ \\
& 2 & $-0.36$ & $-0.13$ & $0.04$ & $1.02$ \\
& 3 & $-0.36$ & $-0.15$ & $0.05$ & $0.89$ \\
& 4 & $-0.38$ & $-0.11$ & $0.02$ & $1.29$ \\
& 5 & $-0.35$ & $-0.16$ & $0.07$ & $0.78$ \\
& 6 & $-0.38$ & $-0.11$ & $0.02$ & $1.31$ \\
& 7 & $-0.37$ & $-0.13$ & $0.03$ & $1.1$ \\     
      \hline
      $101^\circ$& 1 & $-0.19$ & $-0.3$ & $0.05$ & $0.56$ \\
& 2 & $-0.2$ & $-0.29$ & $0.05$ & $0.57$ \\
& 3 & $-0.17$ & $-0.32$ & $0.08$ & $0.49$ \\
& 4 & $-0.25$ & $-0.24$ & $0.01$ & $0.76$ \\
& 5 & $-0.13$ & $-0.36$ & $0.11$ & $0.42$ \\
& 6 & $-0.25$ & $-0.23$ & $0.01$ & $0.78$ \\
& 7 & $-0.21$ & $-0.27$ & $0.04$ & $0.63$       
    \end{tabular}      
  \end{ruledtabular}
  \caption{\label{tab:triang:exchanges}
    Computed exchanges for the seven
    Mg/Ga local environments of Ref.~[\onlinecite{li_crystalline_2017}] as a function
    of bond angle $\theta$. We take the Slater-Koster ratio to be $t_{pf\pi}/t_{pf\sigma} = -0.3$
    and assume $t_{pf\sigma}$ is independent of bond angle and crystal field environment. The
    overall scale is compared between different configurations relative to 
    a reference configuration ($\theta = 99^\circ$, Env. 1)
    }
\end{table}
We now estimate the effects of the Mg/Ga disorder on super-exchange; this includes modification of the crystal field ground state and the change in Yb-O-Yb bond angles. We will not aim at detailed modeling of the local distortions of the YbO\tsub{6} octahedra, instead opting for a rough estimate that captures the qualitative changes that can occur for these kinds of substitutions. To this end, we consider the seven crystal field compositions~\cite{li_crystalline_2017} induced by Mg/Ga disorder (as given in Table~\ref{tab:triangular-cef}) combined with small variations in the bond angles, specifically taking bond angles of $97^\circ$, $99^\circ$ and $101^\circ$. We fix the Slater-Koster ratio to be $\rho = t_{pf\pi}/t_{pf\sigma} = -0.3$ and assume $t_{pf\sigma}$ does not vary strongly with disorder configuration.  As shown in Table~\ref{tab:triang:exchanges}, for all cases we find that $J_{zz}$ is dominant and positive. The transverse coupling $J_{\pm}$ is negative and is the second largest exchange for bond angles $97^{\circ}$ and $99^{\circ}$, while it is competitive with $J_{\pm\pm}$ for $101^\circ$. For all cases $J_{z\pm}$ is relatively small. This is broadly consistent with two reliable features from the literature~\cite{li-2015-rare-earth,li_gapless_2015,paddison_continuous_2016}: that $J_{zz}$ is largest and $J_{\pm}$ is comparable, but smaller, with $J_{\pm} < 0$.  While these exchange ratios vary somewhat between crystal field configurations, they do not change very significantly.  The overall scale, set here to be $J_{zz}$, also varies somewhat between different crystal field configurations, but much more strongly as a function of bond angle. Specifically, as seen in Table~\ref{tab:triang:exchanges}, relative to the $99^\circ$ case, the exchanges change by a factor of 2 when changing the bond angle by $\pm 2^\circ$. We thus tentatively conclude that the primary effect of the Mg/Ga disorder in the super-exchange results is \emph{not} through the variation in doublet composition, but through the variation in bond angle. This introduces strong variations in both the relative importance of the anisotropic couplings, as well as in the overall magnitude of the exchange interactions. This picture of strong exchange disorder in \ybmggao{}, expected from these calculations, supports the rough picture recently put forward in Ref.~[\onlinecite{zhu_disorder-induced_2017}], though it differs in details, and calls into question the interpretation of Refs.~[\onlinecite{paddison_continuous_2016},\onlinecite{shen_evidence_2016}].  Exactly how this appears in the physical properties of \ybmggao{} will depend on the details of the Mg/Ga disorder, for example how spatially correlated it is in the in-plane directions. Disorder with a very short in-plane correlation length (comparable to the rare-earth nearest-neighbor distance) would also introduce a significant lowering of the bond symmetry. In this case, the minimal model of Eq.~(\ref{eq:model}) would be inapplicable; many additional exchange terms (both symmetric and antisymmetric) would be allowed, and thus likely appear, complicating the analysis considerably.

We should also note that the energy scale of the dipolar interactions is non-negligible relative to the exchanges of $O(1\K)$ expected here. As discussed in App.~\ref{app:dipolar:triang}, one expects dipolar contributions to the nearest neighbor exchange of order $\sim 0.1\K$ or so. Since these depend on the $g$-factors they will also be affected by the crystal field disorder, though not (directly) by the changes in the ligand bond angle.

\section{Discussion}
\label{sec:discussion} In this Section, we discuss some limitations of our methodology and explore some more speculative applications to rare-earth pyrochlore oxides of the form \abo{R}{M} and to (potential) Yb-based honeycomb magnets.

First, we comment on the approximations made in the super-exchange calculation of Sec.~\ref{sec:super-exchange} which forms the backbone of this work. Three kinds of approximations were made: in the atomic physics, in the hopping processes and in what processes were included. The most mild are the approximations made in the atomic physics of Yb\tsup{2+}, Yb\tsup{3+} and Yb\tsup{4+}. As discussed in App.~\ref{app:atomic}, we haves not included some of the smaller, more subtle corrections to the intra-shell effective Hamiltonian~\cite{meftah-2013-spectrum}, or included the effect of the crystal field splitting on the excited levels of Yb\tsup{4+} (note that the closed shell Yb\tsup{2+} is trivial). Both of these approximations lead to energy shifts of order a few percent (see App.~\ref{app:atomic}) relative to the bandwidth of the Yb\tsup{4+} states and thus are not expected to be important. This could in principle be remedied by inclusion of corrective terms known in the literature~\cite{meftah-2013-spectrum} and through inclusion of the crystal field explicitly in the atomic calculations. More serious is the uncertainty in the energy costs $U^{\pm}_f$. While these mostly contribute to the overall scale (which we largely ignore), more precise knowledge of the ratio $U^+_f/U^{-}_f$ would be useful in refining these calculations. The uncertainty in the ligand parameters $\DeltaOx$ and $U_p$ also has a similar features. Another relatively mild, but ultimately less controlled, approximation is the use of the two-center Slater-Koster approximation~\cite{slater-1954-simplified}. While this allowed us to reduce the number of free hopping parameters to effectively two, it is unclear how realistic this is. Ideally, the overlaps could be estimated by tight-binding fits to ab-initio band-structure calculations for such rare-earth insulators. Finally, there is the inclusion of only the ligand mediated super-exchange processes. This notably excludes any processes that involve the higher rare-earth orbitals (such as $5d$ or $6s$) or their inter-shell interactions with the $4f$ electrons. While these processes involve intermediate states that are expected to lie higher in energy than the ligands, and are at higher order in perturbation theory, a detailed quantitative estimate of their importance would be helpful in ruling them in or out as significant contributions to the exchange.

Next we comment on the applications of these results more broadly, considering applications to the rare-earth pyrochlore oxides \abo{R}{M} where R is a rare-earth and M a metal ion. The titanate family \abo{R}{Ti}~\cite{gardner-2010-review,gingras-2014-quantum}. The Yb-based compounds \abo{Yb}{M} with M = Ti, Ge, Sn are of particular interest, showing highly unusual dynamic properties~\cite{hallas2016universal}.  While not directly applicable to these compounds due to the presence of two inequivalent exchange paths, it is straightforward in principle to generalize the results of Sec.~\ref{sec:super-exchange} to this case. However, this introduces additional modeling complications, in particular the need to fix two additional hopping parameters (within the Slater-Koster two-center approximation). Given the goal is to determine four exchange parameters, having three tunable hoppings or hopping ratios renders the outcome of the calculation somewhat subjective. However, there are some useful insights that can be gleaned for the idealized case where both exchange paths are equivalent. This corresponds to the case of a perfect cube of oxygens around each Yb ion (with a $\Gamma_7$ doublet ground state) and a bond angle of $\cos^{-1}(-1/3) \sim 109.47^\circ$. As in the case of the $\Gamma_7$ doublet for a $90^\circ$ bond angle, one expects strong suppression of the overall strength thus exchange that is highly sensitive to the details of the calculation, such as the specific values of hopping parameters and the bond-angle (see Secs.~\ref{sec:cubic-limit} and~\ref{sec:general}). While this further exacerbates the difficulties discussed above, it also loosely suggests a rationale for the sensitivity of some members of the \abo{Yb}{M} family to small changes in stoichiometry~\cite{ross2012stuffed} and mild applied pressure~\cite{kermarrec2017ground}.

To conclude, we speculate on some possible interesting Yb-based magnets that have not yet been studied in detail. In particular one is tempted to consider magnets built around the honeycomb of edge-shared octahedra shown in Fig.~\ref{fig:structures}, as has been considered in Kitaev materials such a (Na,Li)\tsub{2}IrO\tsub{3} and RuCl\tsub{3}. One could also consider more complex three-dimensional honeycomb structures as in ($\beta,\gamma$)-Li\tsub{2}IrO\tsub{3}. Indeed the material YbCl\tsub{3} has the needed Yb\tsup{3+} ion and crystallizes in the same structure found in RuCl\tsub{3} (with some monoclinic distortion), with bonds angles of $\sim 97-98^\circ$~\footnote{To compute this angle, we use the atomic positions obtained for YCl$_3$ from Ref.~[\onlinecite{templeton1954crystal}], but the lattice constants and angles determined for YbCl$_3$~\cite{templeton1954crystal}.}. The results for the uniform case discussed in Sec.~\ref{sec:general} apply directly to such unstudied honeycomb magnets; one still expects a very robust region of nearly pure Heisenberg antiferromagnet near the octahedral $\Gamma_6$ limit. While the ground state would be conventional, the appearance of such a nearly isotropic magnet in a rare-earth insulator with very strong spin-orbit coupling would be interesting in and of itself. This would have clear experimental signatures, such as in the appearance of nearly gapless pseudo-Goldstone modes (see for example similar behavior in Sr\tsub{2}IrO\tsub{4}~\cite{rau-2016-arcmp}).  However, as seen in the case of \ybmggao{}, trigonal distortions could (in principle) push the relevant ground doublet composition far from this limit. The existence of strongly Kitaev-like limits in Fig.~\ref{fig:general-colinear} then suggests it may possible, through luck or fine-tuning for these systems, to realize Kitaev's honeycomb model in such a rare-earth insulator.

We thus conclude that rare-earth systems, in particular those based on \rth{Yb} and built from edge-shared octahedra, have the potential to host many different types of anisotropic spin models. In addition to the ``weak" emergent anisotropy found near the ideal octahedral limit, one can also find Kitaev limits and, depending on the lattice, regions when symmetric anisotropies dominate. We hope the varied behavior found in this work and the potential opportunity to explore new realizations of frustrated, anisotropic systems will motivate further studies and development of ytterbium based magnets.

\begin{acknowledgments}
We thank K. A. Ross, L. D. C. Jaubert, G. Chen and M. Mourigal for useful discussions. Research at the Perimeter Institute is supported by the Government of Canada through Innovation, Science and Economic Development Canada and by the Province of Ontario through the Ministry of Research, Innovation and Science.  The work at the University of Waterloo was supported by the NSERC of Canada and the Canada Research Chair program (M.J.P.G., Tier 1).
\end{acknowledgments}

\appendix
\section{Basis conventions}
\label{app:conv}
Here, we outline our basis choices. For the pyrochlore and breathing pyrochlore
lattices we choose the four local frames $(\vhat{x}_i,\vhat{y}_i,\vhat{z}_i)$
\begin{align}
  \vhat{z}_1 &= \frac{1}{\sqrt{3}} \left(+\vhat{x}+\vhat{y}+\vhat{z}\right), &
  \vhat{x}_1 &= \frac{1}{\sqrt{6}} \left(-2\vhat{x}+\vhat{y}+\vhat{z}\right),   \nonumber
  \\
  \vhat{z}_2 &= \frac{1}{\sqrt{3}} \left(+\vhat{x}-\vhat{y}-\vhat{z}\right), &
  \vhat{x}_2 &= \frac{1}{\sqrt{6}} \left(-2\vhat{x}-\vhat{y}-\vhat{z}\right),   \nonumber
  \\
  \vhat{z}_3 &= \frac{1}{\sqrt{3}} \left(-\vhat{x}+\vhat{y}-\vhat{z}\right), &
  \vhat{x}_3 &= \frac{1}{\sqrt{6}} \left(+2\vhat{x}+\vhat{y}-\vhat{z}\right),   \nonumber
  \\  \vhat{z}_4 &= \frac{1}{\sqrt{3}} \left(-\vhat{x}-\vhat{y}+\vhat{z}\right), &
  \vhat{x}_4 &= \frac{1}{\sqrt{6}} \left(+2\vhat{x}-\vhat{y}+\vhat{z}\right),
\end{align}
where $\vhat{y}_i = \vhat{z}_i \times \vhat{x}_i$. The four basis
sites of a tetrahedron, $\vec{r}_i$, are then along the $\vhat{z}_i$ directions,
with $\vec{r}_i = \sqrt{3}\vhat{z}_i/4$.

For the triangular and honeycomb lattices all sites are equivalent
and we define the frame $(\vhat{x}_0,\vhat{y}_0,\vhat{z}_0)$ with
\begin{align}
  \vhat{x}_0 &=  \frac{1}{\sqrt{6}} \left(-2\vhat{x}+\vhat{y}+\vhat{z}\right), & \vhat{z}_0 &=  \frac{1}{\sqrt{3}} \left(+\vhat{x}+\vhat{y}+\vhat{z}\right),
\end{align}
with $\vhat{y}_0 = \vhat{z}_0 \times \vhat{x}_0$. The magnetic ions lie in the
plane perpendicular to $\vhat{z}$ with one of the nearest neighbor bonds
being along $(\vhat{y}-\vhat{z})/\sqrt{2}$.

\section{Atomic physics}
\label{app:atomic}
To obtain the super-exchange contributions that
involve $\f^{12}$ intermediate states, we need to
understand the atomic physics of the (nominal)
Yb$^{4+}$ ion. These energies and states are determined by
the Coulomb interaction and the spin-orbit coupling~\cite{wybourne1965}. Projecting
into the $f$-shell, the Coulomb interaction can be written
\begin{align}
   H_{\rm Coulomb} &= \frac{1}{2}\left(\frac{e^2}{4\pi\epsilon_0}\right) \sum_{i \neq j}\frac{1}{|\vec{r}_i-\vec{r}_j|}, \\
&= \frac{1}{2} \sum_{k=0,2,4,6} a_k F^k \sum_{q=-k}^k \h{O}_{kq} O^{}_{kq},
\end{align} 
where we have defined the numerical coefficients $a_2 = 2/15$, $a_4 = 1/11$, $a_6 = 50/429$.
Microscopically, the Coulomb integrals $F^k$ as defined to be
\begin{equation}
  F^k \equiv \frac{e^2}{4\pi\epsilon_0}  \int^{\infty}_0 dr \int^\infty_0 dr' 
  \left(\frac{r^k_<}{r^{k+1}_>}\right) r^2 (r')^2 R(r)^2 R(r')^2,
\end{equation}
where $r_> = {\rm max}(r,r')$, $r_< = {\rm min}(r,r')$ and $R(r)$ is
the single-particle radial wave-function associated with the
$f$-shell states. The rank-$k$ multipole operators $O_{kq}$ are defined as
\begin{equation}
  O_{kq} \equiv \sqrt{\frac{2l+1}{2k+1}} \sum_{\sigma} \sum_{mm'} (-1)^m 
  \braket{l,-m,l,m'}{k,q} \h{f}_{m\sigma} f^{}_{m'\sigma},
\end{equation}
where $l=3$, $\braket{l,-m,l,m'}{k,q}$ is a Clebsch-Gordan coefficient and
$\h{O}_{kq} = (-1)^q O^{}_{k,-q}$. Note that we have ignored
the $F^0$ Coulomb integral; this is encapsulated in the energy $U^-_f$ defined in the main text.
The spin-orbit coupling takes the form
\begin{align}
  H_{\rm SO} &\equiv \zeta_{\rm SO}\sum_{mm'} \sum_{\sigma\sigma'} 
\left[\vec{\mathcal{L}}_{mm'} \cdot \vec{\mathcal{S}}_{\sigma\sigma'}\right]
\h{f}_{m\sigma} f_{m'\sigma'},
\end{align}
where $\vec{\mathcal{S}} = \vec{\sigma}/2$ (with $\vec{\sigma}$ being the Pauli matrices) and $\vec{\mathcal{L}}$ are 
the angular-momentum matrices for $l=3$.
The total (free-ion) Hamiltonian is then
\begin{equation}
  \label{eq:yb-v-model}
  H_{\rm ion} \equiv H_{\rm Coulomb} + H_{\rm SO}.
\end{equation}

Given the introduction of the solid environment, and the associated screening effects, we will use Coulomb integrals, $F^2$, $F^4$ and $F^6$, and spin-orbit coupling, $\zeta_{\rm SO}$ tailored for Yb$^{4+}$, as determined in Ref.~[\onlinecite{meftah-2013-spectrum}]. These are given by
\begin{align}
  \label{eq:yb-v-param}
  F^2 &= 14.184\ {\rm eV},& 
  F^4 &=  9.846\ {\rm eV},& 
  F^6 &=  6.890\ {\rm eV}, \nonumber \\
  \zeta_{\rm SO} &= 0.380\ {\rm eV}.
\end{align}
The two-hole states of the $f^{12}$ configuration can be constructed
from the basis
\begin{equation}
  \ket{m_1\sigma_1,m_2\sigma_2} \equiv f_{m_1\sigma_1} f_{m_2\sigma_2} \ket{0} 
\end{equation}
where $\ket{0}$ is the filled $f^{14}$ state and $m_1\sigma_1 \neq m_2\sigma_2$.  Since only one of each pair $(m_1\sigma_1,m_2\sigma_2)$ and $(m_2\sigma_2,m_1\sigma_1)$ are independent, we can choose an ordering and thus have ${14 \choose 2} = 91$ states total. We can thus construct the matrix elements of $H_{\rm ion}$ within this subspace directly, given a representation of the fermion operators $f_{m\sigma}$. Diagonalization gives the spectrum shown in Table~\ref{tab:yb-v-spectrum}~\footnote{ This matrix could be further block-diagonalized by using that the total angular momentum $J$ is conserved in a free-ion.  However given the matrix is relatively small as is, we forego such a step.  }. Note that while most of the levels can be identified with the state expected from the $LS$-coupling approximation, there are several levels that exhibit strong mixing.  As expected from Hund's rules, the ground state is primarily composed of states of type $\termsym{1}{H}{6}$ that is $L=5$, $S=1$ and $J=6$.  As noted in Sec.~\ref{sec:super-exchange}, this spectrum spans roughly $\sim 5\ {\rm eV}$ (ignoring the high lying singlet at $\sim 10\ {\rm eV}$). This band-width is comparable to the energy $U^-_f$, thus invalidating the charging approximation~\cite{iwahara-2015-exchange}.

These levels are close to, but not identical, to the experimental results for the spectrum of Yb$^{4+}$~\cite{meftah-2013-spectrum}.  While the agreement can be improved through the inclusion of effective operators that incorporate various correlation effects~\cite{meftah-2013-spectrum}, this level of precision is unimportant for virtual states.  Typically, the differences are likely on the order of $\sim 10-50\ {\rm meV}$, with the highest level having the largest discrepancy of $\sim 150\ {\rm meV}$. Given the size of $U^-_f$ and spread of the levels being $\sim 5-10\ {\rm eV}$, such deviations are on the order of $\sim 1\%-5\%$ and are thus unlikely to be significant for the exchange interactions, at the level of our treatment.

In principle, one could also include the crystal field in the free-ion model of Eq.~(\ref{eq:yb-v-model}). For the trigonal environments considered in this work, this would take the form
\begin{align} H_{\rm CEF} &= C_{2,0} O_{2,0} + C_{4,0} O_{4,0} + C_{6,0} O_{6,0} + C_{4,3} \left(O_{4,+3} - O_{4,-3}\right) \nonumber \\ &+ C_{6,3} \left(O_{6,+3} - O_{6,-3}\right) + C_{6,6} \left(O_{6,+6} + O_{6,-6}\right).
\end{align} where the $C_{kq}$ coefficients can be related to the more common $B_{kq}$ coefficients used when restricting to the single $J$-manifold. However, given that we expect these splittings to be on the order of $\sim 100\ {\rm meV}$ or so, of the same order of the errors in the free-ion levels themselves, we ignore such details in our calculations. This also affords the advantage of parametrizing the effects of the crystal field through the two ground doublet composition parameters $(\eta,\zeta)$ [defined in Eq.~(\ref{eq:cef})] rather than the six $C_{kq}$ variables.

\begin{table}[bp]
  \begin{ruledtabular}    
\begin{tabular}{rrl}
$E\ [{\rm eV}]$ & ${\rm Degeneracy}$ & ${\rm Composition}$\\
\hline
$0.000$ & $13$ & $\termsym{3}{H}{6}$ \\
$0.752$ & $9$ & $\termsym{3}{F}{4}$ \\
$1.187$ & $11$ &$\termsym{3}{H}{5}$ \\
$1.785$ & $9$ & $\termsym{3}{H}{4}$ \\
$2.040$ & $7$ & $\termsym{3}{F}{3}$ \\
$2.094$ & $5$ & $\termsym{3}{F}{2}$ \\
$2.985$ & $9$ & $\termsym{1}{G}{4}$ \\
$3.924$ & $5$ & $\termsym{3}{P}{2}, \termsym{1}{D}{2}$ \\
$4.751$ & $13$ & $\termsym{1}{I}{6}$ \\
$5.023$ & $1$ & $\termsym{3}{P}{0}$ \\
$5.167$ & $3$ & $\termsym{3}{P}{1}$ \\
$5.396$ & $5$ & $\termsym{3}{H}{2}, \termsym{1}{D}{2}$ \\
$10.558$ & $1$ & $\termsym{1}{S}{0}$
\end{tabular}
\end{ruledtabular}
\caption{\label{tab:yb-v-spectrum}
Theoretical spectrum of Yb$^{4+}$ found using the free-ion
model of Eq.~(\ref{eq:yb-v-model}) with parameters given in
Eq.~(\ref{eq:yb-v-param}). The composition indicates the largest
contribution in terms of states constructed through $LS$-coupling. If
there is no dominant component, all significant terms are shown.
}
\end{table}

\section{Super-exchange processes}
\label{app:processes}
 There are twenty-four separate, non-zero contribution to the exchange. We classify these into four types of process separating them based on which intermediate states, i.e. $\f^{12}$ or $\f^{14}$ are involved, and whether one or both ligands, $A$ and $B$, are invoked.  Within each type, only a few are independent; we can obtain many others by interchanging sites $f_1$ and $f_2$ or interchanging the ligands $p_A$ and $p_B$. The first two processes involve only a single ligand at a time (as considered in Refs.~[\onlinecite{onoda-2011-quantum},\onlinecite{rau-2015-magnitude}]), while the final two involve both and are a kind of ring-exchange involving both ligands.

To aid in enumerating the various contributions we will divide $\mathcal{K}$ (see Sec.~\ref{sec:two-ion}) into two pieces, $\mathcal{K}_1$ that couples to $O_1 \tilde{O}_2$ and $\mathcal{K}_2$ that couples to $\tilde{O}_1 O_2$, keeping in mind that the bond symmetries force $\mathcal{K}_1 = \mathcal{K}_2$ in the final result [see Eq.~(\ref{eq:o-defn})].

\newcommand{\con}[4]{f^{#1}_1 p_A^{#2} p^{#3}_B f_2^{#4}}
\subsection{Process 1}
We first consider a class of process that involves both the $\f^{14}$ and $\f^{12}$ states. There are four such processes, but only one is elementary, the remaining three can be obtained by swapping $f_1$ and $f_2$ and $p_A$ and $p_B$. This process is
\begin{align}
 \mathcal{P}_1^{(1)}  &: \con{13}{6}{6}{13} \rightarrow
  \con{14}{5}{6}{13} \rightarrow 
  \con{14}{6}{6}{12} \rightarrow \nonumber \\
 &\con{14}{5}{6}{13} \rightarrow 
  \con{13}{6}{6}{13}  .
\end{align}
It contributes to the exchange the operator
\begin{align}
  &-\sum_{\alpha\beta\mu\nu}\sum_{\alpha'\beta'\mu'\nu'}
  \frac{
  t^{\alpha\beta}_{1A} \left[\h{t}_{2A}\right]^{\nu\mu}
    {t}_{2A}^{\alpha'\beta'}
    \left[\h{t}_{1A}\right]^{\nu'\mu'}
  }{(\Uf +\DeltaOx)^2(U^+_f+U^-_f)} \nonumber \\
  & P 
    \left(\h{p}_{A\nu'} f^{}_{1\mu'}\right)
    \left(\h{f}_{2\alpha'} p^{}_{A\beta'}\right) Q_2
    \left(\h{p}_{A\nu} f^{}_{2\mu}\right)
    \left(\h{f}_{1\alpha} p^{}_{A\beta}\right)
  P  .
\end{align}
Using the fact that the ligand part is given by
\begin{equation}
  P^{}_p\h{p}_{A\nu'} p^{}_{A\beta'} \h{p}_{A\nu} p^{}_{A\beta}P_p = \delta_{\beta\nu}\delta_{\beta'\nu'}.
\end{equation}
The contribution of this process is then
\begin{align}
  &+\sum_{\alpha\beta\mu\nu}
  \frac{
  \left[t^{}_{1A} \h{t}_{2A}\right]^{\alpha\nu}
  \left[t^{}_{2A} \h{t}_{1A}\right]^{\mu\beta}
  }{(\Uf +\DeltaOx)^2(U^+_f+U^-_f)} 
    \left(P^{}_1\h{f}_{1\alpha} f^{}_{1\beta}P^{}_1\right)
    \left(P^{}_2\h{f}_{2\mu} Q^{}_2 f^{}_{2\nu} P^{}_2\right). \nonumber
\end{align}
This is thus a contribution to $\mathcal{K}_1$. Swapping $A$ and $B$ gives an additional contribution to $\mathcal{K}_1$, while swapping $1$ and $2$ gives a contribution to $\mathcal{K}_2$. These contributions are identical so, in total, one has a final contribution to $\mathcal{K}$ given by
\begin{align}
  \mathcal{K}^{\alpha\beta\mu\nu} :&   
\sum_{\lambda=A,B} \frac{
  T_{\lambda}^{\alpha\nu}
  \left[\h{T}_{\lambda}\right]^{\mu\beta}
  }{(\Uf +\DeltaOx)^2(U^+_f+U^-_f)} .
\end{align}
Note that these contributions are Hermitian on their own, since $\trp{T}_{\lambda} = T^{}_{\lambda}$, $\h{[O^{\alpha\beta}]} = O^{\beta\alpha}$ and $\h{[\tilde{O}^{\alpha\beta}]} = \tilde{O}^{\beta\alpha}$.
\subsection{Process 2}
We next consider a process involving only the $\f^{14}$ states and
only a single ligand at a time. There are eight exchange paths in
total. However there are only two elementary processes;
we consider each in turn. 
\subsubsection{Process $\mathcal{P}^{(1)}_2$}
The first process is given by
\begin{align}
 \mathcal{P}_2^{(1)} &: 
  \con{13}{6}{6}{13} \rightarrow
  \con{14}{5}{6}{13} \rightarrow 
  \con{14}{4}{6}{14} \rightarrow \nonumber \\
 &\con{13}{5}{6}{14} \rightarrow 
  \con{13}{6}{6}{13}  .
\end{align}
This $\mathcal{P}^{(1)}_2$ process contributes
\begin{align}
  &-\sum_{\alpha\beta\mu\nu}\sum_{\alpha'\beta'\mu'\nu'}
  \frac{
  t^{\alpha\beta}_{1A} t_{2A}^{\mu\nu}
    \left[\h{t}_{1A}\right]^{\beta'\alpha'}
    \left[\h{t}_{2A}\right]^{\nu'\mu'}
  }{(\Uf +\DeltaOx)^2(2(U^+_f+\DeltaOx)+U_p)} \times \nonumber \\
  & P 
    \left(\h{p}_{A\nu'} f^{}_{2\mu'}\right)
    \left(\h{p}_{A\beta'} f^{}_{1\alpha'}\right)
    \left(\h{f}_{2\mu} p^{}_{A\nu}\right)
    \left(\h{f}_{1\alpha} p^{}_{A\beta}\right)
  P  .
\end{align}
The ligand part is trivial and yields
\begin{equation}
  P^{}_p\h{p}_{A\nu'} \h{p}_{A\beta'} p^{}_{A\nu} p^{}_{A\beta}P_p = \delta_{\beta'\nu}\delta_{\beta\nu'}-\delta_{\beta\beta'}\delta_{\nu\nu'}.
\end{equation}
The second piece gives terms that involve $t_{1A}^{} \h{t}_{1A}$ and
such; these do not contribute for \rth{Yb} once projected into the
ground state manifold~\cite{rau-2015-magnitude}. The final relevant pieces are thus
\begin{align}
  &+\sum_{\alpha\beta\mu\nu}
  \frac{
  \left[t^{}_{1A} \h{t}_{2A}\right]^{\alpha\nu}
  \left[t^{}_{2A} \h{t}_{1A}\right]^{\mu\beta}
  \left(P^{}_1 \h{f}_{1\alpha}f^{}_{1\beta}
    P^{}_1\right)
  \left(P^{}_2
    \h{f}_{2\mu} 
    f^{}_{2\nu}    
    P^{}_2\right) 
  }{(\Uf +\DeltaOx)^2(2(U^+_f+\DeltaOx)+U_p)} .
\end{align}
\subsubsection{Process $\mathcal{P}^{(2)}_2$}
The second process is given by
\begin{align}
 \mathcal{P}_2^{(2)} &: 
  \con{13}{6}{6}{13} \rightarrow
  \con{14}{5}{6}{13} \rightarrow 
  \con{14}{4}{6}{14} \rightarrow \nonumber \\
 &\con{14}{5}{6}{13} \rightarrow 
  \con{13}{6}{6}{13}  .
\end{align}
This $\mathcal{P}^{(2)}_2$ process contributes
\begin{align}
  &-\sum_{\alpha\beta\mu\nu}\sum_{\alpha'\beta'\mu'\nu'}
  \frac{
  t^{\alpha\beta}_{1A} t_{2A}^{\mu\nu}
    \left[\h{t}_{2A}\right]^{\beta'\alpha'}
    \left[\h{t}_{1A}\right]^{\nu'\mu'}
  }{(\Uf +\DeltaOx)^2(2(U^+_f+\DeltaOx)+U_p)} \times \nonumber \\
  & P 
    \left(\h{p}_{A\nu'} f^{}_{1\mu'}\right)
    \left(\h{p}_{A\beta'} f^{}_{2\alpha'}\right)
    \left(\h{f}_{2\mu} p^{}_{A\nu}\right)
    \left(\h{f}_{1\alpha} p^{}_{A\beta}\right)
  P  .
\end{align}
The ligand part is again trivial and yields
\begin{equation}
  P^{}_p\h{p}_{A\nu'} \h{p}_{A\beta'} p^{}_{A\nu} p^{}_{A\beta}P_p = \delta_{\beta'\nu}\delta_{\beta\nu'}-\delta_{\beta\beta'}\delta_{\nu\nu'}.
\end{equation}
The first piece gives terms that involve $t_{1A}^{} \h{t}_{1A}$ and
such; again these do not contribute for \rth{Yb} once projected into the
ground state manifold. The relevant pieces are thus
\begin{align}
  &+\sum_{\alpha\beta\mu\nu}
  \frac{
  \left[t^{}_{1A} \h{t}_{2A}\right]^{\alpha\nu}
  \left[t^{}_{2A} \h{t}_{1A}\right]^{\mu\beta}
  \left(P^{}_1 \h{f}_{1\alpha}f^{}_{1\beta}
    P^{}_1\right)
  \left(P^{}_2
    \h{f}_{2\mu} 
    f^{}_{2\nu}    
    P^{}_2\right) 
  }{(\Uf +\DeltaOx)^2(2(U^+_f+\DeltaOx)+U_p)} .
\end{align}
We thus see that extra sign from the ligand part is compensated by the
sign from rearranging the $f$ operators. This thus gives the same
contribution as the first process ($\mathcal{P}^{(1)}_2$).
\subsubsection{Total}
Both processes, $\mathcal{P}^{(1)}_2$ and $\mathcal{P}^{(2)}_2$,  contribute to $\mathcal{I}$.  Putting this all
together, swapping $f_1$ and $f_2$ as well as $p_A$ and $p_B$, we find
the contribution from the eight processes of type 2 are given by
\begin{align}
  \mathcal{I}^{\alpha\beta\mu\nu} &:
& 4 \sum_{\lambda=A,B} \frac{
  T_{\lambda}^{\alpha\nu}
  \left[\h{T}_{\lambda}\right]^{\mu\beta}
  }{(\Uf +\DeltaOx)^2(2(U^+_f+\DeltaOx)+U_p)}     .
\end{align}
\subsection{Process 3}
We next consider the simpler of the two ring-exchange processes.
This involves both $\f^{14}$ and $\f^{12}$ states. There
is only a single elementary process given as
\begin{align}
 \mathcal{P}_3^{(1)} &: 
  \con{13}{6}{6}{13} \rightarrow
  \con{14}{5}{6}{13} \rightarrow 
  \con{14}{6}{6}{12} \rightarrow \nonumber \\
 &\con{14}{6}{5}{13} \rightarrow 
  \con{13}{6}{6}{13}  .
\end{align}
This $\mathcal{P}^{(1)}_3$ process contributes
\begin{align}
  &-\sum_{\alpha\beta\mu\nu}\sum_{\alpha'\beta'\mu'\nu'}
  \frac{
    t_{1A}^{\alpha\beta}
    \left[\h{t}_{2A}\right]^{\nu\mu}
    t_{2B}^{\alpha'\beta'}
    \left[\h{t}_{1B}\right]^{\nu'\mu'}
  }{(\Uf +\DeltaOx)^2(U^+_f+U^-_f)} \times \nonumber \\
  & P 
    \left(\h{p}_{B\nu'} f^{}_{1\mu'}\right)
    \left(\h{f}_{2\alpha'} p^{}_{B\beta'}\right) Q^{}_2
    \left(\h{p}_{A\nu} f^{}_{2\mu}\right)
    \left(\h{f}_{1\alpha} p^{}_{A\beta}\right)
  P  .
\end{align}
The ligand part is (again) trivial and yields
\begin{equation}
  P^{}_p\h{p}_{B\nu'} {p}^{}_{B\beta'} \h{p}_{A\nu} p^{}_{A\beta}P_p = \delta_{\beta'\nu'}\delta_{\beta\nu}.
\end{equation}
We thus have
\begin{align}
  &+\sum_{\alpha\beta\mu\nu}
  \frac{
    \left[t^{}_{1A} \h{t}_{2A}\right]^{\alpha\nu}
    \left[t_{2B}^{} \h{t}_{1B}\right]^{\mu\beta}
    \left(P^{}_1 \h{f}_{1\alpha}f^{}_{1\beta} P^{}_1\right)
    \left(P^{}_2 \h{f}_{2\mu} Q^{}_2 f^{}_{2\nu}P^{}_2\right)
  }{(\Uf +\DeltaOx)^2(U^+_f+U^-_f)} .
\end{align}
This is a contribution to $\mathcal{K}_1$. Swapping $A$ and $B$ generates another
contribution to $\mathcal{K}_1$ while swapping $1$ and $2$
generates (identical) contributions to $\mathcal{K}_2$. One has
thus has a net contribution to $\mathcal{K}$ given by
\begin{align}
  \mathcal{K}^{\alpha\beta\mu\nu} &:
 \sum_{\lambda=A,B} \frac{
  T_{\lambda}^{\alpha\nu}
  \left[\h{T}_{\cb{\lambda}}\right]^{\mu\beta}
  }{(\Uf +\DeltaOx)^2(U^+_f + U^-_f)},
\end{align}
where $\cb{\lambda}$ is the other ligand; i.e. $\cb{A} = B$ and $\cb{B}=A$. The sum
over the two ligands renders these combined contributions Hermitian.
\subsection{Process 4}
The final type of process involves only $f^{14}$ states and both
ligands. There are eight different paths; two of these are
independent.
\subsubsection{Process $P^{(1)}_4$}
The first process is given as
\begin{align}
 \mathcal{P}_4^{(1)} &: 
  \con{13}{6}{6}{13} \rightarrow
  \con{14}{5}{6}{13} \rightarrow 
  \con{14}{5}{5}{14} \rightarrow \nonumber \\
 &\con{14}{6}{5}{13} \rightarrow 
  \con{13}{6}{6}{13}  .
\end{align}
This gives the contribution
\begin{align}
  &-\sum_{\alpha\beta\mu\nu}\sum_{\alpha'\beta'\mu'\nu'}
  \frac{
    t_{1A}^{\alpha\beta}
    t_{2B}^{\mu\nu}
    \left[\h{t}_{2A}\right]^{\beta'\alpha'}
    \left[\h{t}_{1B}\right]^{\nu'\mu'}
  }{2(\Uf +\DeltaOx)^3} \times \nonumber \\
  & P 
    \left(\h{p}_{B\nu'} f^{}_{1\mu'}\right)
    \left(\h{p}_{A\beta'} f^{}_{2\alpha'}\right)
    \left(\h{f}_{2\mu} p^{}_{B\nu}\right)
    \left(\h{f}_{1\alpha} p^{}_{A\beta}\right)
  P  .
\end{align}
The ligand part is trivial and yields
\begin{equation}
  P^{}_p\h{p}_{B\nu'} \h{p}_{A\beta'} {p}^{}_{B\nu} p^{}_{A\beta}P_p = -\delta_{\beta\beta'}\delta_{\nu\nu'}.
\end{equation}
This then leads to the contribution
\begin{align}
  &+\sum_{\alpha\beta\mu\nu}\sum_{\alpha'\mu'}
  \frac{
    \left[t^{}_{1A} \h{t}_{2A}\right]^{\alpha\nu}
    \left[t^{}_{2B} \h{t}_{1B}\right]^{\mu\beta}
  }{2(\Uf +\DeltaOx)^3} 
    \left( P^{}_1
    \h{f}_{1\alpha} 
    f^{}_{1\beta} P^{}_1\right)
    \left(P^{}_2
    \h{f}_{2\mu} 
    f^{}_{2\nu}P^{}_2\right).
\end{align}
\subsubsection{Process $P^{(2)}_4$}
Next, we move on to the second process. This has the form
\begin{align}
 \mathcal{P}_4^{(2)} &: 
  \con{13}{6}{6}{13} \rightarrow
  \con{14}{5}{6}{13} \rightarrow 
  \con{14}{5}{5}{14} \rightarrow \nonumber \\
 &\con{13}{5}{6}{14} \rightarrow 
  \con{13}{6}{6}{13}  .
\end{align}
This contributes
\begin{align}
  &-\sum_{\alpha\beta\mu\nu}\sum_{\alpha'\beta'\mu'\nu'}
  \frac{
    t_{1A}^{\alpha\beta}
    t_{2B}^{\mu\nu}
    \left[\h{t}_{1B}\right]^{\beta'\alpha'}
    \left[\h{t}_{2A}\right]^{\nu'\mu'}
  }{2(\Uf +\DeltaOx)^3} \times \nonumber \\
  & P 
    \left(\h{p}_{A\nu'} f^{}_{2\mu'}\right)
    \left(\h{p}_{B\beta'} f^{}_{1\alpha'}\right)
    \left(\h{f}_{2\mu} p^{}_{B\nu}\right)
    \left(\h{f}_{1\alpha} p^{}_{A\beta}\right)
  P  .
\end{align}
The ligand part is trivial and yields
\begin{equation}
  P^{}_p\h{p}_{A\nu'} \h{p}_{B\beta'} {p}^{}_{B\nu} p^{}_{A\beta}P_p = \delta_{\beta'\nu}\delta_{\beta\nu'}.
\end{equation}
This leads to
\begin{align}
  &+\sum_{\alpha\beta\mu\nu}
  \frac{
    \left[t^{}_{1A} \h{t}_{2A}\right]^{\alpha\nu}
    \left[t^{}_{2B} \h{t}_{1B}\right]^{\mu\beta}
  }{2(\Uf +\DeltaOx)^3}
    \left(P^{}_1 \h{f}_{1\alpha}
    f^{}_{1\beta} P^{}_1\right)
    \left(P^{}_2 \h{f}_{2\mu} 
    f^{}_{2\nu}P^{}_2\right),
\end{align}
identical to the previous process, $\mathcal{P}^{(1)}_4$.
\subsubsection{Total}
We thus see that the total contribution  of all the $P^{(n)}_4$ processes is to $\mathcal{I}$, and
is given by
\begin{align}
  \mathcal{I}^{\alpha\beta\mu\nu} &:
4 \sum_{\lambda=A,B} \frac{
  T_{\lambda}^{\alpha\nu}
  \left[\h{T}_{\cb{\lambda}}\right]^{\mu\beta}
  }{2(\Uf +\DeltaOx)^3}  .
\end{align}

\section{Dipolar interactions}
\label{app:dipolar}
Due to the small energy scale associated with rare-earth
super-exchange, we must also consider direct magnetostatic
dipole-dipole interactions between the ytterbium ions when comparing
directly to materials. The full exchange interactions will be the sum
of these dipolar terms and the super-exchange interactions derived in
Sec.~\ref{sec:super-exchange}. The dipolar interactions take the form
\begin{equation}
  H_{\rm MDD} = \frac{\mu_0}{4\pi}\sum_{i < j} \frac{1}{|\vec{r}_{ij}|^3}\left[
    \vec{\mu}_i \cdot \vec{\mu}_j - 3(\vhat{r}_{ij} \cdot \vec{\mu}_i)(\vhat{r}_{ij} \cdot \vec{\mu}_j)
    \right],
\end{equation}
where $\vec{r}_{ij} \equiv \vec{r}_i -\vec{r}_j$ with $\vec{r}_i$ the
position of rare-earth ion $i$
and the magnetic moment $\vec{\mu}_i$ is defined in terms of the
pseudo-spins $\vec{S}_i$ in Eq.~(\ref{eq:moment}). For simplicity, we
consider only the nearest-neighbor part of the dipolar interaction.
Since this depends on the lattice geometry we discuss the two cases of
experimental interest, the \spinel{A}{Yb}{X} spinels and \ybmggao{},
separately.
\subsection{Breathing pyrochlore and spinels}
\label{app:dipolar:spinels}
Given the local axes appropriate for the  breathing pyrochlore
and the spinels, we can explicitly
compute the form of the nearest neighbor part of $H_{\rm MDD}$ and map it to
the local exchanges $J_{zz}$, $J_{\pm}$, $J_{\pm\pm}$ and $J_{z\pm}$. One finds
in the notation of Sec.~\ref{sec:two-ion}~\cite{wong2013xy}
\begin{align}
  J_{zz} &= +\frac{5}{3} \mathcal{D} g_z^2, &
  J_{\pm} &= -\frac{1}{12} \mathcal{D} g_{\pm}^2, \nonumber \\
  J_{\pm\pm} &= +\frac{7}{12} \mathcal{D} g_{\pm}^2, &
  J_{z\pm} &= -\frac{1}{3\sqrt{2}} \mathcal{D} g_{\pm}g_z, 
\end{align}
where $\mathcal{D} \equiv \mu_0 \mu_B^2/(4\pi r_{\rm nn})^3$ with
$r_{\rm nn}$ being the nearest-neighbor distance. In \byzo{},
the nearest-neighbor distance is $\sim 3.3\AA$~\cite{rau-2016-byzo}
yielding $\mathcal{D} \sim 0.0173 \K$. Combined with the $g$-factors
$g_z \sim -2.73$ and $g_{\pm} \sim -2.3$ yields the dipolar
contribution to the exchanges (in the
appropriate global basis)
\begin{align}
  J_{\rm d} &\sim -0.006\meV, &
  K_{\rm d} &\sim +0.015\meV, \nonumber \\
  \Gamma_{\rm d} &\sim -0.014\meV, &
  D_{\rm d} &\sim -0.002\meV.    
\end{align}
These represent small perturbations to the dominant Heisenberg ($J \sim 0.6\meV$) and \ac{DM} ($|D| \sim 0.18\meV$) exchanges. Since for the parameters the ground state is a (symmetry protected) $E$-doublet with a large gap to the higher excited states~\cite{rau-2016-byzo}, the small symmetric anisotropies can be ignored.

In the \spinel{A}{Yb}{S} spinels, the
nearest-neighbor distance is roughly $\sim 3.9\AA$, while in the
\spinel{A}{Yb}{Se} spinels, it is closer to
$4.1\AA$~\cite{higo-2016-frustrated}. For these distances, one has the
energy scale $\mathcal{D} \sim 0.01\K$ which must be combined with
typical $g$ factors of $g_z \sim -3.6$ and $g_{\pm} \sim -3.2$. In the
global basis appropriate for the spinels, this yields a very similar
result to \byzo{}, with the dipolar contribution to the
nearest-neighbor exchanges being roughly
\begin{align}
  J_{\rm d} &\sim -0.006\meV, &
  K_{\rm d} &\sim +0.015\meV, \nonumber \\
  \Gamma_{\rm d} &\sim -0.014\meV, &
  D_{\rm d} &\sim -0.02\meV.    
\end{align}
Note that the near equality of these contributions for the breathing pyrochlore and spinels is an accident; both the moment size (encoded in the $g$-factors) and the nearest-neighbor distances are different, but they nearly compensate each other.

For the spinels, in contrast to the breathing pyrochlore case, since we are proximate to a phase boundary which is controlled by the symmetric anisotropies, we must treat these somewhat carefully.  To estimate the importance of these corrections in the spinels, we first set the overall scale of the exchanges using the Curie-Weiss temperature~\cite{ross-2011-quantum}. This yields, roughly, that $J \equiv J_{\rm s} + J_{\rm d} \sim 0.3\meV - 0.35\meV$, depending on the spinel under consideration ($J_{\rm s}$ is the super-exchange contribution and $J_{\rm d}$ the dipolar part). With this energy scale set, the super-exchange contribution to the symmetric anisotropies are thus $K_{\rm s} \sim -0.01\meV$ and $\Gamma_{\rm s} \sim -0.006\meV$ and \ac{DM} contribution is $D_{\rm s} \sim -0.1 \meV$.  In total, one then has $D/J \sim -0.3$, $K/J = +0.0125$ and $\Gamma/J = -0.05$. We thus see that $K+\Gamma = K_{\rm s} + K_{\rm d} + \Gamma_{\rm s} +\Gamma_{\rm d} < 0$ and thus we still expect selection of a state from the $\Gamma_5$ manifold (as opposed to an \ac{SFM} state). In summary, while small, the dipolar corrections are not ignorable since the sub-dominant $K$, $\Gamma$ interactions control the ground state selection for the relevant exchange regime. However, since $K_{\rm d} +\Gamma_{\rm d} \lesssim K_{\rm s} + \Gamma_{\rm s}$ this does not affect the conclusions of Sec.~\ref{sec:spinels}.

\subsection{Triangular}
\label{app:dipolar:triang}
For the triangular lattice of \ybmggao{}, the structure of the dipolar
interaction is simpler due to the frames being the same from
site to site. In the notation of Sec.~\ref{sec:two-ion} one finds that
\begin{align}
  \label{eq:dipolar:triang}
  J_{d,zz} &= +\mathcal{D} g_z^2, &
  J_{d,\pm} &= +\frac{1}{4} \mathcal{D} g_{\pm}^2, \nonumber \\
  J_{d,\pm\pm} &= +\frac{3}{4} \mathcal{D} g_{\pm}^2, &
  J_{d,z\pm} &= 0,
\end{align}
where $\mathcal{D} \equiv \mu_0 \mu_B^2/(4\pi r_{\rm nn})^3$ with
$r_{\rm nn}$ being the nearest neighbor distance. For \ybmggao{}
the nearest-neighbor distance is $r_{\rm nn} \sim 3.4\AA$~\cite{li-2015-rare-earth} and thus
one has $\mathcal{D} \sim 0.0158\K$. Taking typical $g$-factors to
be $g_z \sim 3.72$ and $g_{\pm} \sim -3.06$~\cite{li-2015-rare-earth}, one arrives at
\begin{align}
  J_{d,zz} &= +0.22\K, &
  J_{d,\pm} &= +0.04\K, \nonumber \\
  J_{d,\pm\pm} &= +0.11\K, &
  J_{d,z\pm} &= +0.00 \K.
\end{align}
The fitted exchange parameters (adapted to our notation) of
Ref.~[\onlinecite{li-2015-rare-earth}] are similar in scale to those
listed here, with $J_{zz} \sim |J_{\pm}| \sim 1\K $ with
$|J_{d\pm\pm}|\sim 0.155 \K$ and $J_{z\pm} \sim 0$. We thus see that to make
any meaningful comparison of the super-exchange result to the fitted
exchanges, this dipolar part \emph{must} be properly subtracted from
the fitted result. Due to the dependence on the $g$-factors, these
exchanges will also be sensitive to the crystal field disorder
found in Ref.~[\onlinecite{li_crystalline_2017}], and discussed in Sec.~\ref{sec:triangular}. Using the
appropriate crystal field parameters~\cite{li_crystalline_2017}, one
finds the dipolar contributions listed in
Table~\ref{tab:dipolar:triang}. The most significant variation is in
$J_{d,zz}$, given the $g$-factor $g_z$ also experiences the largest
changes as a function of environment.

\begin{table}[hb]
  \begin{ruledtabular}
    \begin{tabular}{ccccc}
      Env. & $J^{}_{d,zz}$ [K] &$J^{}_{d,\pm}$ [K] &$J^{}_{d,\pm\pm}$ [K] &$J^{}_{d,z\pm}$ [K] \\
      \hline
      1 & $0.22$ & $0.04$ & $0.12$ & $0.00$ \\
      2 & $0.21$ & $0.04$ & $0.12$ & $0.00$ \\
      3 & $0.24$ & $0.04$ & $0.12$ & $0.00$ \\
      4 & $0.16$ & $0.04$ & $0.13$ & $0.00$ \\
      5 & $0.26$ & $0.04$ & $0.11$ & $0.00$ \\
      6 & $0.16$ & $0.04$ & $0.13$ & $0.00$ \\
      7 & $0.20$ & $0.04$ & $0.13$ & $0.00$ 
    \end{tabular}
  \end{ruledtabular}
  \caption{\label{tab:dipolar:triang}
    Dipolar contributions to the nearest-neighbor anisotropic
exchange, as given in Eq.~(\ref{eq:dipolar:triang}), in \ybmggao{}
using the seven Mg/Ga disorder induced crystal field environments of
Ref.~[\onlinecite{li_crystalline_2017}].
  }
\end{table}

\bibliography{draft}

\end{document}